\documentclass[aps,prx,groupedaddress,reprint,showpacs,longbibliography]{revtex4-1}
\usepackage{graphicx}
\usepackage{tabularx}
\usepackage{dcolumn}
\usepackage{bm}
\usepackage{amsmath}
\usepackage{amssymb}
\usepackage{txfonts}
\usepackage{url}
\usepackage{subfig}
\usepackage{float}
\newcommand{\Lot}{L1$_2$}
\newcommand{\Dzt}{D0$_{22}$}
\newcommand{\aTm}{$\bar{T}_m$}

\begin{document}

\title{
  Order--Disorder Competitive Cooperation in Equiatomic 3d--Transition--Metal Quaternary Alloys: Phase Stability and Electronic Structure
}

\author{Hiroshi Mizuseki}
\affiliation{Korea Institute of Science and Technology (KIST), Seoul 02792, Republic of Korea,}
\author{Ryoji Sahara}
\affiliation{National Institute for Materials Science (NIMS), Tsukuba 305-0047, Japan,}
\author{Kenta Hongo$^{*}$}
\affiliation{
  Research Center for Advanced Computing Infrastructure, JAIST,
  Asahidai 1-1, Nomi, Ishikawa 923-1292, Japan
}
  
\date{\today}
\begin{abstract}

We use high-throughput first-principles sampling to investigate competitive factors
that determine the crystal structure of high-entropy alloys (HEAs) 
and the energetics dependence of the stable phase on the atomic configuration
of fully ordered \Lot{},
\Dzt{}, and random solid solution (RSS) phases 
of equiatomic quaternary alloys comprising four 
of the six constituent elements (Cr, Mn, Fe, Co, Ni, and Cu).
Considering the configurational entropy, 
we demonstrate that valence electron concentration (VEC)
and temperature 
are crucial to determine the phase stability
of HEAs at finite temperatures, wherein
the ordered phases are energetically more favorable than RSS phases.
Some \Dzt{} phases with high VEC
are energetically more stable than \Lot{} phases,
though both phases are metastable.
Further, we explore magnetic configurations to identify the origin of the enthalpy term.
The calculations reveal that ordered phases comprising 
antiferromagnetic atoms surrounded
by ferromagnetic atoms are energetically stable.
The quantitative structure--property relationship is also discussed.
\end{abstract}
\maketitle

\section{Introduction}
\label{sec.intro}

High-entropy alloys (HEAs)~\cite{HEAbook,2020GEO} 
have been gaining increasing attention because of their 
novel material properties that cater to 
a wide range of applications.
Although HEAs were originally applied to 
structural materials that required both strength and
ductility,~\cite{2016VAR,2017VAR,2019WEI,2020MAR,2021HAN}
their recent applications have been extended to other functional materials
such as catalysis,~\cite{2021SUN} superconductors,~\cite{2021MIZ}
radiation resistance materials~\cite{2019EL-}, and
magnetic materials.~\cite{2015KOZ,2021MA}
These extensions of their applications represent a new direction of materials design
based on the fundamental and key concept of ``high-entropy.''
Along with this new direction,
the comprehensive understanding of quantitative structure--property relationship (QSPR)
is crucial for exploiting their potential applications.
However, its most fundamental information---microscopic/atomistic crystal structures
of HEAs---remains unknown~\cite{2021SAV};
this implies that the atomistic characterization of HEA structures is challenging
for both ``experiments'' and ``theories/computations,'' 
in terms of ``spatial resolutions''~\cite{2018YAO} and ``computational costs for
vast amounts of possible atomic configurations,''~\cite{2018LED,2021UTI} respectively.

It is considered that HEAs have random solid solutions (RSSs)
instead of ordered phases.
Although disordered face-/body-centered cubic (FCC/BCC) phases (A1/A2) have been
reported as single HEA phases,~\cite{2020CHA}
ordered FCC/BCC (\Lot{}/B2) and hexagonal close-packed (HCP, A3) have
also been found.~\cite{2014ZHA,2017MIR,2017ROG}
Although there are a few examples of ordered phases, 
perfect randomness in some types of HEAs
has been excluded in several studies. 
For example, short range order (SRO) in HEAs
has been reported~\cite{2018DIN,2019LIb,2020ZHAb};
however, the existence of long-range order (LRO) is rarely discussed
and remains a controversial topic.~\cite{2012LUC, 2015NIU}
The most famous ordered phase in HEAs is
CrFeCoNi, with the \Lot{} structure,~\cite{2015NIU}
wherein the antiferromagnetic Cr atoms occupy the ordered lattice sites 
surrounded by randomly distributed ferromagnetic Ni, Fe, and Co atoms.
This can be interpreted as a spin-driven stabilization of the atomic configuration, which indicates 
that enthalpy enhancement by magnetization 
cooperates {\it partial} randomness
and then overcomes the entropy
because of {\it perfect} randomness.
This was theoretically predicted~\cite{2014MID,2015NIU,2017FUK}
and verified experimentally only for the ``surface''
using scanning transmission electron microscopy (STEM)~\cite{2015NIU};
experimental verifications for the ``bulk'' have not been feasible 
because the constituent atoms are adjoining elements in the periodic table and have similar chemical properties and comparable radii, which 
makes them indistinguishable when using spectroscopic methods.~\cite{2018YAO}
Thus, the question arises if the spin-driven stabilization
is unique for CrFeCoNi or if it generally holds for other 
quaternary alloys.
To address this issue from a microscopic perspective, instead of experiments,
first-principles approaches are the most relevant
for identifying atomistic structures of HEAs
and elucidating their QSPR~\cite{2018CHO,2019RIC,2019YIN,2021WIT},
though the previous studies dealt with only single composites.

We systematically focus on 
3d-transition-metal-based quaternary alloys
with Cr, Mn, Fe, Co, Ni, and Cu as the constituent elements; 
a total of $15 ~(= \binom{6}{4} =: \frac{6!}{4!2!})$ different types of
composites are considered~(TABLE~\ref{table.main}).
These elements are most commonly used magnetic HEA constituents ~\cite{2017MIR},
and all quaternary alloy combinations that results from them can be speculated to have the FCC phase~\cite{2017DIA}.
Therefore, they are expected to be disordered/RSSs 
as their atomic sizes and the nature of chemical bonds formed by them are similar.
Our first-principles study revealed 
that differences in their magnetic properties
with respect to CrFeCoNi lead to different magnetic orderings:
our target phases were not restricted to only RSSs and \Lot{} ordered phases; they extended to the \Dzt{} ordered phases as well (Figure~\ref{figure.str}),
motivated by 
well-known studies on order--disorder phase competition
in ordered alloys~\cite{1983BIE,1996CAB,2018STO,2019HE,2019ZHO,2019ZHA}.
The \Lot{} and \Dzt{} phases are
ordered crystal structures for 25:75 at.\% FCC alloys.
As prototype systems for ordered phases, 
their phase competition has been 
well investigated for conventional alloys
in terms of valence electron concentration (VEC).~\cite{1983BIE,1996CAB}
For example, the \Lot{} phase of intermetallic compounds is observed
for VEC $\ge$ 7.5.
However, to the best of our knowledge, there is no previously reported study
on their counterpart for HEAs; 
in HEAs, VEC values are known to correlate well with structure types
(FCC/BCC structures prefer higher/lower
VEC values).~\cite{2011GUO1,2011GUO2,2015TIA,2017LEO}
Thus, we analyze the RSS--\Lot{}--\Dzt{} thermodynamical competition
in the 3d-transition-metal quaternary system in terms of VEC.

We performed high-throughput first-principles sampling from
four types of \Lot{} and \Dzt{} phases to understand the RSS--\Lot{}--\Dzt{} competition and their QSPR,
and from one type of RSS phase for 15 different composites.
We found that composites with lower VEC values 
tend to have ordered phases over wide temperature ranges; however,
for higher VEC values, RSSs emerge as the temperature increases.
In addition, we discuss 
the distribution of the magnetic moment at the first nearest neighbor site
and the constituent elements in our HEAs. 
We believe that phase stability is realized 
by order--disorder ``competitive cooperation''
rather than by the ``commonly conceived'' order--disorder competition:
the former is caused by the cooperation between a ``magnetically ordered 
atomic configuration composed of one element gaining enthalpy''
and ``fully disordered/random configuration composed of the other three elements
gaining entropy,'' which we propose in this study;
the latter is the realization of either 
``fully ordered configuration'' or ``fully random configuration.''
Their difference is not sufficiently appreciated in the HEA community.

The paper is organized as follows:
Computational details in the present study are given in ``METHODS''.
In ``RESULTS'', we start with our main finding
``Order--Disorder Competitive Cooperation'', followed by
``Energetics'' and ``Magnetic ordering'',
which can be explained by ``Quantitative Structure--Property Relationship''.
``SUMMARY AND FUTURE PROSPECTS'' summarize our findings and
their significance towards further investigation in HEAs.

\section{Methods}
\label{sec.methods}
\subsection{HEA structural models}
\label{subsec.model}
HEAs have been regarded as solid solutions,
wherein every site in the crystal structure is randomly occupied by the constituent elements.
However, it is difficult to reproduce such a random configuration with a small simulation cell size.
Therefore, finding an appropriate method to mimic the configuration of HEAs
under a limited periodic condition is crucial for theoretical investigations.
This problem can be solved
by evaluating the correlation functions of the atomic configurations in HEAs. 
For instance, Zunger et al. proposed 
the concept of special quasirandom structures (SQSs) ~\cite{1990ZUN}, 
which are supercell approximations for a disordered system 
optimized to mimic the random local atomic environment. 
Therefore, the SQS approach is widely applied to HEA research 
because it provides a good approximation for the RSS phase. 
However, it is difficult to discuss the energetics and phase stability of HEAs in detail 
using only one configuration obtained from the SQS model. 
It has previously been reported that applying several SQS configurations of HEAs 
leads to non-negligible energy fluctuations (0.01--0.02 eV/atom).~\cite{HEAbook,2018NIU}
We performed high-throughput first-principles sampling 
to provide insights regarding the phase stability of HEAs with high accuracy and reliability for overcoming the disadvantage of using a single configuration in the SQS approach.

Two-, three-, and four-body clusters were selected 
for a correlation function, with the Warren--Cowley short-range order (SRO) parameter 
evaluated by homogeneous and heterogeneous elements~\cite{1960COW,1965COW}.
The range of the SRO parameters was restricted to 
the fourth nearest-neighbor site for two-body clusters and 
to the third nearest-neighbor site for three- and four-body clusters 
to reduce the computational time. 
For the RSS phase, the configurations of four elements 
were explored to realize a better correlation function 
using the Metropolis algorithm~\cite{1953MET}
with a wide range of fictitious temperatures. 
For both the ordered and RSS phases, 
four elements were considered for evaluating the correlation function.
In the \Lot{} ordered phase, 
one element was fixed at the cubic corner (CC) sites in the FCC lattice, 
while the other three elements in the quaternary system 
occupied the face-centered sites to realize an ideal \Lot{} ordered phase
with a good correlation function. 
The \Dzt{} phase was also constructed in a similar manner. 
A wide range of occupancy rates was used
to elucidate the influence of 
the number of first nearest-neighbor sites 
on the formation energy. 
Accordingly, a large number of atomic configurations
were randomly generated for RSS structures (Fig.~\ref{figure.ed}).
Our structural models of the quaternary alloys were
described by $2~\!\times~\!2~\!\times~\!2$ supercells
with 32 sites or $4~\!\times~\!4~\!\times~\!4$ supercells with 256 sites,
depending on the computed quantities;
the equiatomic system comprises 
8 atoms in the $2~\!\times~\!2~\!\times~\!2$ supercell and 
64 atoms in the $4~\!\times~\!4~\!\times~\!4$ supercell for each element. 
The crystal structures were drawn using the VESTA software~\cite{2011MOM}.

 \subsection{High-throughput first-principles sampling}
\label{subsec.dft}
The free energy of formation (Figure~\ref{figure.fe}) 
and magnetic moment (Figure~\ref{figure.mm}) 
for each phase were obtained 
based on 100 configurations using a $2~\!\times~\!2~\!\times~\!2$ supercell.
Figure S-2 shows the formation free energies obtained 
using 100 configurations. 
The dependences of the formation energy on the occupancy rate 
(Figure~\ref{figure.ed}) 
and the bond-length distributions 
(Figure~\ref{figure.rdf}) 
for the CrFeCoNi quaternary alloy were 
obtained based on 1,500 and 100 configurations, respectively,
using a $4~\!\times~\!4~\!\times~\!4$ supercell.

\par
We carried out high-throughput spin-polarized first-principles simulations
based on DFT~\cite{1965KOH} 
using the Vienna ab initio simulation package (VASP) by inputing all generated HEA models above~\cite{1996KRE1,1996KRE2}. 
The PBEsol functional~\cite{2008PER} was selected 
for the electronic structure calculations and geometry relaxation. 
Projector-augmented wave (PAW)~\cite{1994BLO,1999KRE} potentials
were used to consider the interactions between the ion cores and valence electrons.
The Brillouin zone was described by a set of k-points in a $3~\!\times~\!3~\!\times~\!3$ grid mesh
using the Monkhorst--Pack method.~\cite{1976MON}
For structural optimization,
the energy convergence criterion for electronic iterations
was set at $10^{-2}$ eV/\AA{}. 
The lattice parameter
was set at 3.495 \AA{} during structural optimization to reduce the calculation time.
As an initial condition, Fe, Co, Ni, and Cu were set as ferromagnetic, whereas Cr
was set as antiferromagnetic. 
For systems without Cr, 
Mn was set as antiferromagnetic; otherwise, 
Mn was set as ferromagnetic as an initial condition 
because its magnetism is affected by Cr.

\subsection{Formation energy and formation free energy}
\label{subsec.fe}
To discuss phase stability, the formation energy $E_f(\rm{HEA})$ is defined as
\begin{equation}
  \label{eq:fe}
  E_f({\rm{HEA}}) = E({\rm{HEA}}) - \sum_i x_i E(X_i).
\end{equation}
Here, $E(\rm{HEA})$ denotes the total energy per atom of the HEA
with the alloying elements; $X_i$ and $x_i$ represent the fractions of the alloying elements;
and $E(X_i)$ denote the energies per atom of the alloying elements $X_i$ in their ground state structures, i.e., 
BCC Cr, $I\bar{4}3m$ Mn, 
BCC Fe, HCP Co, FCC Ni, and FCC Cu,
which are estimated by referring to the Open Quantum Materials Database~\cite{2013SAA,2015KIR}. 
A negative value of the formation energy implies that the HEA is stable,
whereas a positive value implies that it is less stable than the ground states of the pure alloying elements.

Temperature is one of the most important factors
that governs the phase stability of materials because most materials are used in a finite temperature range
instead of being used at absolute zero.~\cite{2017ZHO}.
Here, the Helmholtz free energy is defined as 
\begin{eqnarray}
  \label{eq:ffe}
  \nonumber
  F(V,T) &=& E_{\mathrm{el}}(V)
  + E_{\mathrm{vib}}(V, T) \\ 
  &-& TS_{\mathrm{ele}}(V, T)
  - TS_{\mathrm{vib}}(V, T)
  - TS_{\mathrm{config}}(V, T),
\end{eqnarray}
where $E_{\mathrm{el}}(V)$ denotes the internal energy,
and the phonon vibration contribution includes
the lattice vibration energy, i.e., $E_{\mathrm{vib}}(V, T)$,
and lattice vibration entropy, i.e., $S_{\mathrm{vib}}(V, T)$. 
 Further, $S_{\mathrm{config}}(V, T)$ represents the atomic configurational entropy, 
and the last term $S_{\mathrm{ele}}(V, T)$
denotes the electronic-scale entropy contribution,
which includes thermal excitation and spin polarization.
In this study, contributions from the vibrational 
and electronic entropy terms are not included 
owing to the limitations of the calculations.~\cite{2019NIU}
Further, the vibrational contributions for both the ordered and RSS phases were estimated and found to be 
comparable~\cite{1998WAL}.
The contributions of the electronic
and magnetic terms are smaller than
that of the configurational term~\cite{2015MA}. 
$S_{\mathrm{config}}$ in the configurational entropy term 
is given by $S_{\mathrm{config}} = R\sum_i c_i \ln c_i$,
where $c_i$ denotes the ratio between the number of atoms 
of a disordered component and the total number of disordered atoms, and
$R$ represents the gas constant. 
In this study, we introduced an equiatomic quaternary system and 
calculated $S_{\mathrm{config}} = R \ln(3)\times 0.75 = 0.824 R$ 
for the ordered phases and 
$S_{\mathrm{config}} = R \ln(4) = 1.386 R$
for the RSS phase. 
In other words, 
the RSS phase has a greater configurational entropy contribution.
Therefore, the free energy of formation is calculated by
\begin{equation}
  \label{eq:hfe}
  F_f({\rm{HEA}}) = E({\rm{HEA}}) - TS_{\mathrm{config}}.
\end{equation}

\section{Results}
\label{sec.resultsDiscussion}

\subsection{Order--Disorder Competitive Cooperation}
\label{subsec.comp}
One of the main objectives of this study is to investigate whether an ordered phase (OP$_0$) is more stabilized than the corresponding
RSS for each studied quaternary alloy.
That is, we intended to determine the ``crossover'' temperature  ($T_c$) 
at which the phase transition between the OP$_0$ and RSS occurs
for each composite of the quaternary alloy.
Therefore, it is important to consider the configurational entropy term 
to discuss phase stability in HEAs in the finite temperature region. 
It is imperative to quantitatively analyze the influences of VEC and temperature 
on the stable HEA crystal structures
for investigating the order--disorder competitive cooperation and phase stability.
This is because the enthalpy term defined at the absolute zero temperature correlates with VEC, whereas the entropy effect is temperature dependent; the most stable phase is
determined by free energy, i.e., the sum of the enthalpy and entropy terms.
Since the direct elucidation of QSPR is difficult, 
material features (or descriptors) such as VEC~\cite{2011GUO1,2011GUO2}
are ``inserted'' between ``structure'' and ``property'';
then, two step analyses are conducted based on
``structure--feature'' and ``feature--property'' relationships.
Guo et al.~\cite{2011GUO1,2011GUO2} reported
several HEA systems wherein the
single FCC phase was stable in the high VEC region (VEC $> 8.0$),
the single BCC phase was stable in the low VEC region (VEC $< 6.87$),
and the mixed FCC and BCC phases appeared
in the intermediate region ($6.87 < \rm{VEC} < 8.0$).
BCC and FCC enhance the strength and 
ductility, respectively~\cite{HEAbook}.
The average VEC values of magnetic HEAs are greater than 7.5, and therefore, they possess the FCC structure. Hence, 
this study does not focus on the lattice type of magnetic HEAs,
but on their magnetic orderings.
We investigate the suitability of VEC to
identify the orderings, which
would be the first step toward successful designing HEAs.

\par
\begin{figure}[htbp]
  \begin{center}
    \includegraphics[width=1.0\linewidth]{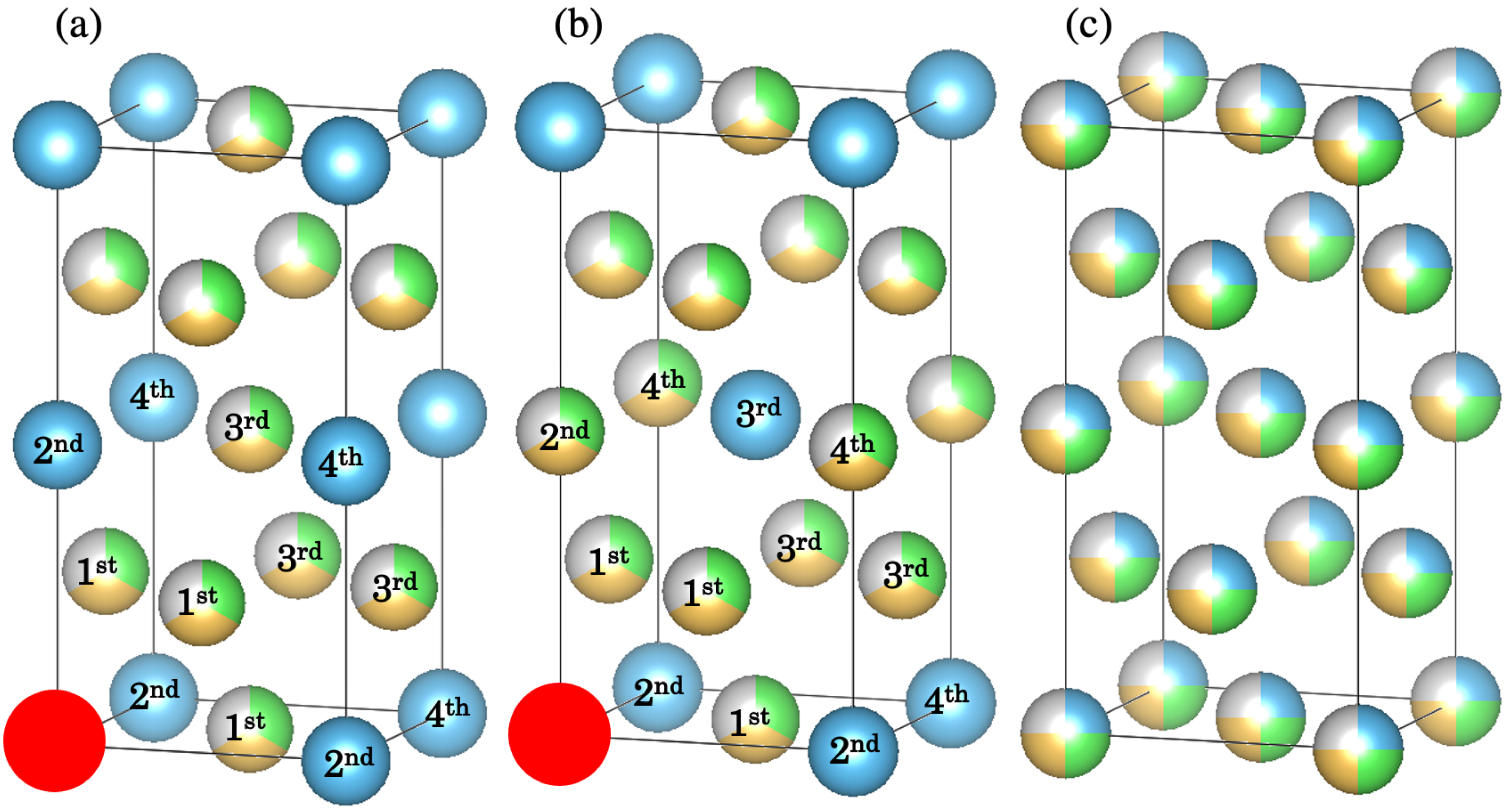}
    \caption{
      Crystal structures of (a) \Lot{} (Cu$_3$Au-type, space group $Pm\bar{3}m$, no. 221),
      (b) \Dzt{} (Al$_3$Ti-type, space group $I4/mmm$, no. 139), and (c) RSS for quaternary alloys. 
      Labels indicate the first, second, third, and fourth nearest neighbor sites from the red site 
      located at the cubic corner.
    }
    \label{figure.str}
  \end{center}  
\end{figure}

TABLE~\ref{table.main} lists 
the quaternary alloy composites investigated in this study (Composite), 
average VECs (VEC$_{\mathrm{ave}}$), 
average melting temperature of the four constituent elements in the alloy (\aTm),
ordered phase with the lowest energy at 0 K (OP$_0$), 
temperature at which stability is attained for each phase ($T_s(X)$ ($X =$ \Lot{}, \Dzt{}, RSS)),
``crossover'' temperature between the ordered phase OP$_0$ and RSS ($T_c$), 
and the corresponding free energy of formation at $T_c$ ($F_\mathrm{c}$).
The the most stable solid phase (MSSP) of each composite
below \aTm{} and the corresponding temperature range ($\Delta T_s$)
are also listed in the table. 
The free energy of formation considering the configurational entropy contribution 
is elucidated in Eq.~\eqref{eq:ffe}.
To estimate estimating the free energy of formation,
we can consider the ordered phases as a 75 at.\% equiatomic ternary alloy 
for the configurational entropy term
because one of the elements is completely located at a specific site
(Fig.~\ref{figure.str} (a) or (b)).
\begin{table*}
 \begin{center}
   \caption{
     Temperature dependence of the most stable structures for 
     15 cases of equiatomic quaternary systems (column name: ``Composite'') 
     investigated in the present study.
     VEC$_\mathrm{ave}$ corresponds to the average valence electron concentration of the composite, 
     where Cr, Mn, Fe, Co, Ni, and Cu have VEC values of 6, 7, 8, 9, 10, and 11, respectively. 
     \aTm{} represents the average melting temperature of the four elements that comprise the composite.
     OP$_0$ represents the ordered phase possessing the lowest energy at 0 K. 
     For example, Cr-\Lot{} indicates the ordered phase 
     wherein the ordered lattice positions are occupied by Cr. 
     (NB: This does not necessarily mean the ordered phase has 
     a lower energy than the RSS.)
     $T_s(X)$ ($X$ = \Lot{}, \Dzt{}, RSS) indicates the temperature 
     at which phase $X$ is stabilized, i.e., 
     its free energy of formation becomes zero. 
     $T_c$ represents a ``crossover'' temperature between the ordered phase OP$_0$ and RSS. 
     $F_c$ (eV/atom) denotes the free energy of formation at $T_c$. 
     $\Delta T_s$ represents the temperature range up to $T_m$, 
     wherein either one of the ordered phases or the RSS phase
     is realized as the most stable solid phase (MSSP).
     All temperatures are listed in Kelvin (K).
   }
   \label{table.main}
   \begin{tabular}{lccccccccc}
                      &                   &      &        &\multicolumn{3}{c}{$T_s$}               &        &              &                     \\ \cline{5-7}
       Composite(\#) & VEC$_\mathrm{ave}$ & \aTm & OP$_0$  & \Lot & \Dzt & RSS & $T_c$  & $F_\mathrm{c}$ & MSSP ($\Delta T_s$ ) \\ 
       \hline
        CrMnFeCo(1)  & 7.50 & 1820 & Cr-\Lot  &   380 &  683 &  638 & 1016 & -0.045  & \Lot{} (380 $\sim$ 1016) \\  
                     &      &      &          &       &      &      &      &         & RSS (1016 $\sim$ \aTm) \\  
        CrMnFeNi(2)  & 7.75 & 1810 & Cr-\Lot  &   448 &  737 &  657 &  963 & -0.037  & \Lot{} (448 $\sim$ 963) \\  
                     &      &      &          &       &      &      &      &         & RSS (963 $\sim$ \aTm) \\  
        CrMnCoNi(3)  & 8.00 & 1799 & Cr-\Lot  &   282 &  553 &  574 & 1003 & -0.051  & \Lot{} (282 $\sim$ 1003) \\  
                     &      &      &          &       &      &      &      &         & RSS (1003 $\sim$ \aTm) \\  
        CrMnFeCu(4)  & 8.00 & 1717 & Cr-\Lot  &  2594 & 2755 & 1820 &  687 &  0.135  & unstable  \\ 
        CrFeCoNi(5)  & 8.25 & 1872 & Cr-\Lot  &     0 &  127 &  494 & 1333 & -0.100  & \Lot{} (0 $\sim$ 1333) \\  
                     &      &      &          &       &      &      &      &         & RSS (1333 $\sim$ \aTm) \\  
        CrMnCoCu(6)  & 8.25 & 1706 & Cr-\Lot  &  2251 & 2425 & 1649 &  768 &  0.105  & RSS{} (1649 $\sim$ \aTm)  \\
        CrFeCoCu(7)  & 8.50 & 1779 & Cr-\Lot  &  1913 & 1917 & 1469 &  819 &  0.078  & RSS{} (1469 $\sim$ \aTm) \\ 
        CrMnNiCu(8)  & 8.50 & 1696 & Cr-\Lot  &  1881 & 1950 & 1398 &  691 &  0.085  & RSS{} (1398 $\sim$ \aTm)  \\ 
        MnFeCoNi(9)  & 8.50 & 1706 & Mn-\Lot  &     0 &  128 &  333 &  970 & -0.076  & \Lot{} (0 $\sim$ 970)  \\ 
                     &      &      &          &       &      &      &      &         & RSS{} (970 $\sim$ \aTm) \\  
        CrFeNiCu(10) & 8.75 & 1769 & Cr-\Dzt  &  1786 & 1546 & 1223 &  750 &  0.074  & RSS{} (1223 $\sim$ \aTm)\\ 
        MnFeCoCu(11) & 8.75 & 1614 & Mn-\Lot  &  1597 & 1665 & 1218 &  663 &  0.066  & RSS{} (1218 $\sim$ \aTm)\\ 
        CrCoNiCu(12) & 9.00 & 1758 & Cr-\Dzt  &  1711 & 1594 & 1117 &  417 &  0.092  & RSS{} (1117 $\sim$ \aTm)  \\
        MnFeNiCu(13) & 9.00 & 1604 & Mn-\Dzt  &  1395 & 1205 &  956 &  592 &  0.057  & RSS{} (956 $\sim$ \aTm)\\ 
        MnCoNiCu(14) & 9.25 & 1593 & Mn-\Dzt  &  1497 & 1204 &  981 &  653 &  0.060  & RSS{} (981 $\sim$ \aTm)\\ 
        FeCoNiCu(15) & 9.50 & 1666 & Fe-\Lot  &   421 &  474 &  469 &  539 & -0.008  & \Lot{} (421 $\sim$539)\\  
                     &      &      &          &       &      &      &      &         & RSS (539 $\sim$ \aTm) \\  
       \hline
     \end{tabular}
 \end{center}
\end{table*}

Consequently, one element of the ordered phases can be ignored 
for the number of the elements considering the configurational entropy term. 
Alternatively, the RSS phases need to be treated as a quaternary alloy. 
Therefore, the contribution of the configurational entropy term for the RSS phase
is larger than that for the ordered phases. 
Accordingly, the slope of the free energy of formation for the RSS phase is steeper than 
that for the ordered phases. 
The ordered phases consequently appear gradually as the temperature increases, and this is followed by the appearance of the RSS phases at higher temperatures.
The above discussion presupposes thermal equilibrium.
Note that ordered phases that thermally equilibrate at higher temperatures 
can appear even at ambient temperatures by non-equilibrium processing,
i.e., rapid solidification and cooling.~\cite{2013YEH}

\begin{figure}[htbp]
  \begin{center}
    \includegraphics[width=1.0\linewidth]{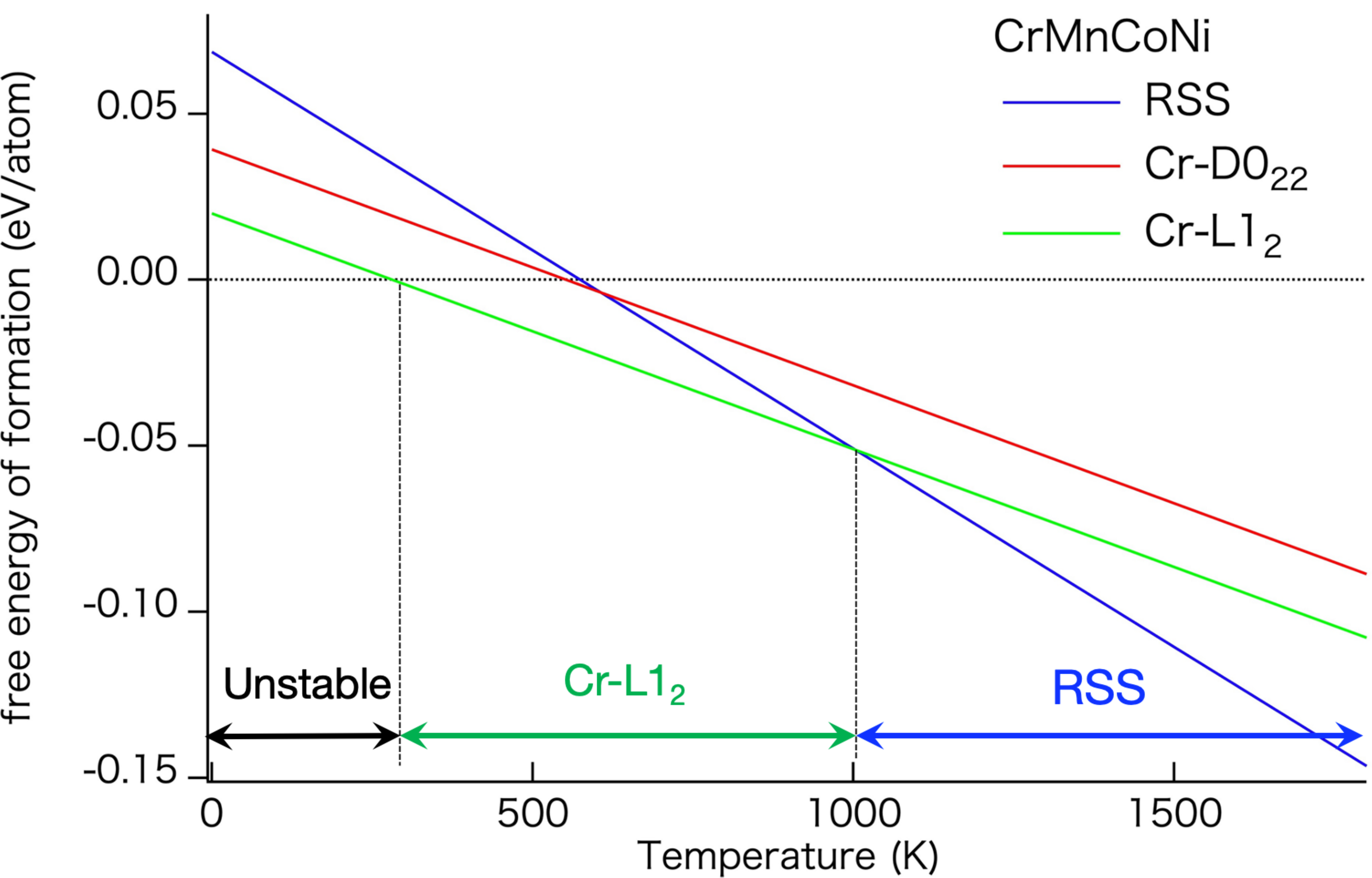}        
    \caption{
      Free energy of formation of CrMnCoNi quaternary alloy
      as a function of temperature up to its melting temperature (1799 K).
      Blue, red, and green lines indicate
      the RSS, Cr-\Dzt{}, and Cr-\Lot{} ordered phases, respectively.
    }
    \label{figure.crsovr}
  \end{center}  
\end{figure}

Figure~\ref{figure.crsovr} shows the free energy of formation of the 
CrMnCoNi quaternary alloy as a function of temperature. 
The temperature dependence of 
the free energy of formation of the other alloys, 
including the CrMnCoNi alloy, is shown in Fig.~S-1 (supplementary information).
For the both ordered phases, Cr-\Lot{} and Cr-\Dzt{}
are the most stable phase of each ordered phase (Fig.~\ref{figure.fe} (3)).
Therefore, we compare the free energies of formation of the Cr-\Lot{} and 
Cr-\Dzt{} ordered phases and the RSS phase up to \aTm (1799 K).
The temperature dependence can be seen as follows:
\begin{enumerate}
\item Within the low-temperature region (up to 282 K), 
  all ordered and RSS phases exhibit a positive free energy of formation, 
  and they are less stable than
  the corresponding ground state of each pure element. 
\item At 282--1003 K, Cr-\Lot{} shows a negative free energy of formation, 
  which is lower than those of the Cr-\Dzt{} and RSS phases. 
  Therefore, Cr-\Lot{} is the most stable phase in this temperature region. 
\item Beyond 1003 K, the RSS phase shows
  a lower free energy of formation than Cr-\Lot{} and 
  becomes a stable phase in the high-temperature region.
\end{enumerate}
Meanwhile, the Cr-\Dzt{} phase does not
appear as the most stable phase in the entire temperature range. 
Consequently, in the case of the CrMnCoNi quaternary alloy 
(VEC$_{\mathrm{ave}} = 8.00$), the \Lot{} crystal structure appears 
as an ordered phase in a specific temperature region.

\par
Further, as shown in Fig.~S-1 and TABLE~\ref{table.main},
the CrMnFeCo (1; VEC$_{\mathrm{ave}} = 7.5$),
 CrMnFeNi (2; VEC$_{\mathrm{ave}} = 7.75$),
 CrFeCoNi (5; VEC$_{\mathrm{ave}} = 8.25$),
 MnFeCoNi (9; VEC$_{\mathrm{ave}} = 8.5$), and 
 FeCoNiCu (15; VEC$_{\mathrm{ave}} = 9.5$) 
quaternary alloys exhibit 
the same phase transition between the \Lot{} ordered phase and RSS phase 
at a specific temperature. 
However, the CrFeCoNi (5; VEC$_{\mathrm{ave}} = 8.25$) and 
 MnFeCoNi (9; VEC$_{\mathrm{ave}} = 8.5$) quaternary alloys 
exhibited the \Lot{} structure at 0 K.
Furthermore,
 the CrMnCoCu (6; VEC$_{\mathrm{ave}} =$ 8.25),
 CrFeCoCu (7; VEC$_{\mathrm{ave}} =$ 8.5),
 CrMnNiCu (8; VEC$_{\mathrm{ave}} =$ 8.5),
 CrFeNiCu (10; VEC$_{\mathrm{ave}} =$ 8.75),
 MnFeCoCu (11; VEC$_{\mathrm{ave}} =$ 8.75),
 CrCoNiCu (12; VEC$_{\mathrm{ave}} =$ 9.0),
 MnFeNiCu (13; VEC$_{\mathrm{ave}} =$ 9.0), and
 MnCoNiCu (14; VEC$_{\mathrm{ave}} =$ 9.25),
quaternary alloys show the RSS phase only
below the melting temperature 
because of the small difference in the free energy of formation 
at 0 K. (Fig.~S1 (6)-(8), and (9)-(14))
The CrMnFeCu (4; VEC$_{\mathrm{ave}} =$ 8.0)
quaternary alloy is unstable in the entire temperature range.

\par
\begin{figure}[htbp]
  \begin{center}
    \includegraphics[width=1.0\linewidth]{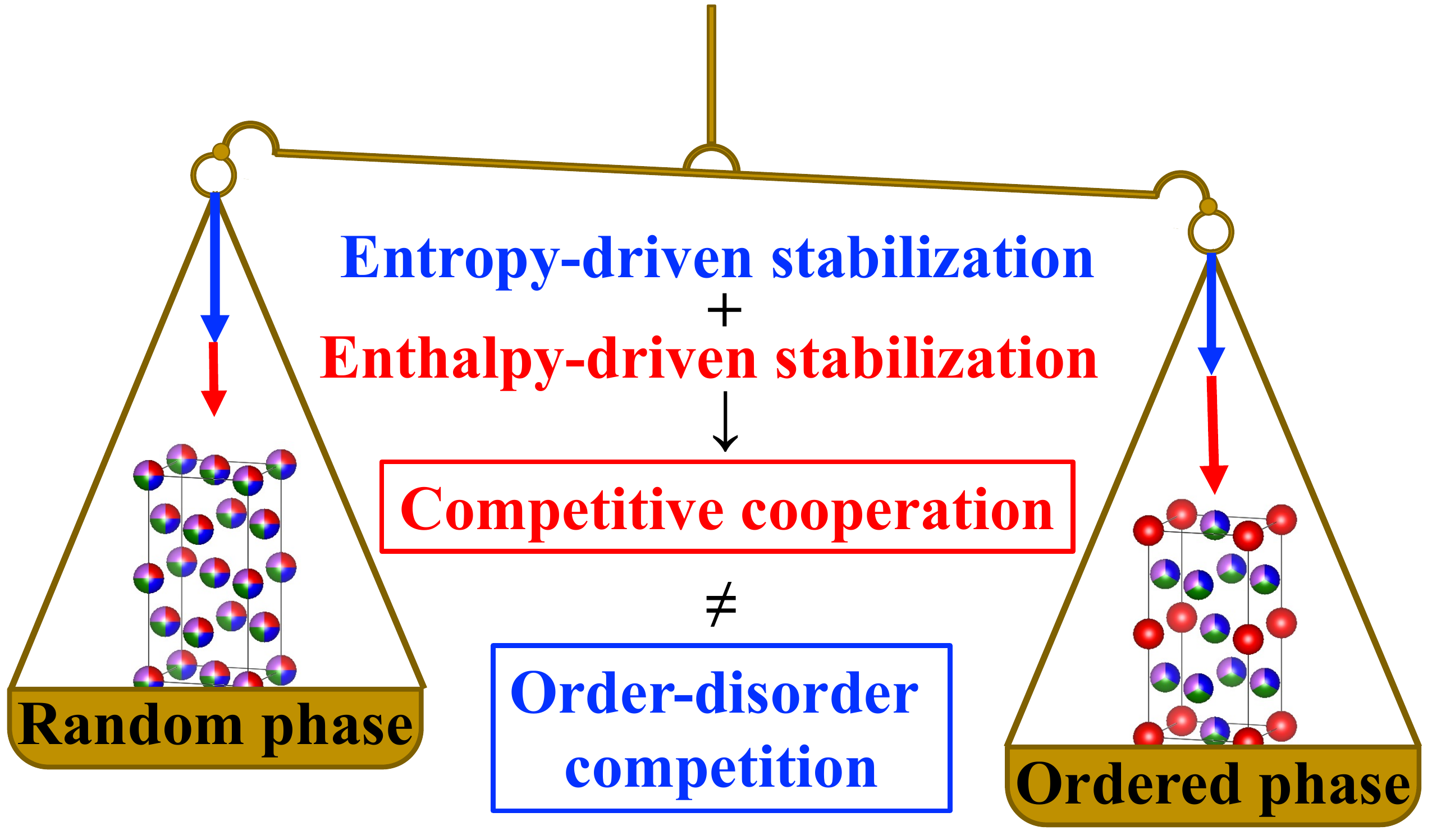}
    \caption{Phase stability of 3d-transition metal
      high-entropy alloys (HEAs) is realized by 
      an order–-disorder ``competitive cooperation,''
      rather than a usual order–-disorder competition.
      The former was newly proposed in this study, 
      while the latter has already been accepted in the HEA community.~\cite{2015NIU}}
    \label{figure.toc}
  \end{center}
\end{figure}

The results of this study are summarized as follows:
The alloys without Cu as a constituent prefer the 
\Lot{} ordered phases for small VEC values ($\le 8.25$).
The Mn-based \Lot{} of MnFeCoNi (9) and the Fe-based \Lot{} of FeCoNiCu (15) are also stable under a specific temperature.
The free energy of formation of the \Dzt{} ordered phases are lower than \Lot{} ordered phases
for large VEC values ($\ge 8.50$), 
except for MnFeCoCu (11) and FeCoNiCu (15).
However, \Dzt{} ordered phases do not appear as the most stable structure, 
because there is no condition that the free energy of formation is negative 
and smaller than the RSS in the entire temperature range.
In contrast, alloys with large VEC values and Cu as a constituent exhibit the RSS phases.
Finally, we may conclude that neither entropy nor enthalpy is dominant
in stabilizing HEAs, but both {\it cooperatively} stabilize their structures (see Figure~\ref{figure.toc})
depending on temperature and other factors such as VEC.

\subsection{Energetics}
\label{subsec.ene}
\begin{figure*}[htbp]
  \begin{center}
    \includegraphics[width=1.0\linewidth]{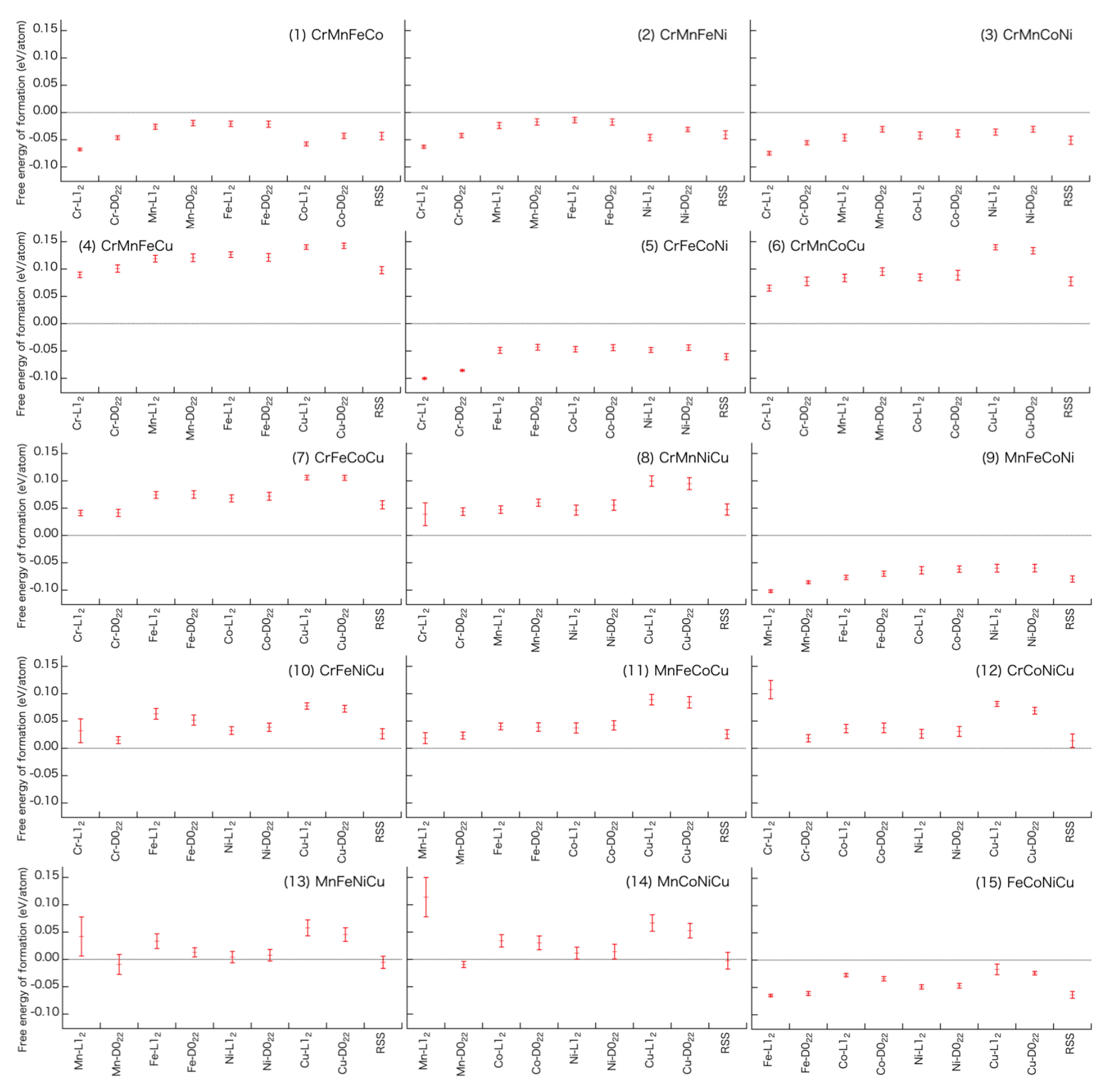}
    \caption{
      Free energy of formation at 1000 K for 15 cases of quaternary alloys 
      listed in Table~\ref{table.main}.
      Error bars indicate the standard deviations (SD) for each phase.
      The original raw data for each phase are shown in Fig.~S-2 (Supplementary Information).
    }
    \label{figure.fe}
  \end{center}  
\end{figure*}

A clear understanding of the energetics of each phase is important
to determine the free energy of formation at a finite temperature for discussing the phase stability of HEAs.
Figure~\ref{figure.fe} shows the average free energies of formation of 100 different configurations, 
with the standard deviation at 1000 K for all 15 quaternary alloys, 
each of which has eight types of ordered phases and the RSS phase. 
Ordered phases with the lowest energy at 0 K (OP$_0$) 
for each quaternary alloy are listed in Table~\ref{table.main}.
For all quaternary alloys (Fig.~S-1), the free energy of formation 
of at least one ordered phase is lower than that of the RSS phase at 0 K.
Interestingly, the \Dzt{} ordered phases 
exhibit a lower free energy of formation than the \Lot{} ordered phases
in the high VEC region in some alloys.
However, because the free energy of formation of 
\Dzt{} ordered phases is not negative below the RSS value,
the \Dzt{} phases are not stable phase in the entire temperature range.

\par
As shown in Fig.~\ref{figure.fe}, CrFeNiCu (10; VEC$_{\mathrm{ave}} = 8.75$), 
CrCoNiCu (12; VEC$_{\mathrm{ave}} = 9.0$), 
MnFeNiCu (13; VEC$_{\mathrm{ave}} = 9.0$), 
and MnCoNiCu (14; VEC$_{\mathrm{ave}} = 9.25$) exhibit
 \Dzt{} phases with lower energies than that of the \Lot{} phases.
These alloys demonstrate that the free energy of \Lot{} increases remarkably, 
as a result, these free energies of formation show a positive value at 1000K.
This result means that these ordered structures are thermodynamically unstable, 
however, we compare the free energies of \Lot{} and \Dzt{} to get insight of a competitive cooperation.
Similar competitive behavior between the phase stability of \Lot{} and \Dzt{} 
for a wide range of VECs was previously reported for Ni$_3$V, Pd$_3$V, and Pt$_3$V, 
which are fundamental binary alloys, in contrast with HEAs~\cite{1996CAB}.
The VEC at which the crossover of the energy difference between the \Lot{} and \Dzt{} phases
occurs in this study (VEC$_{\mathrm{ave}} = 8.5$) is comparable to the value reported 
for the previously studied binary system~\cite{1996CAB}.
One key result obtained from Fig.~\ref{figure.fe} 
is that the systems comprising Cu have high free energies of formation. 
Moreover, quaternary alloys in which Cu forms the ordered phase 
entail higher free energies of formation of the ordered phase than those of the RSS phase. 
We attribute the high energies of formation associated with Cu 
to the positive formation energies exhibited by the least energetic structures of the binary compounds 
consisting of Cu and 3d transition metals (Cr, Mn, Fe, Co, and Ni)~\cite{2015TRO},~\cite{2021KUC},~\cite{2005TAK}.
The present results are consistent with previous studies because 
as the number of bonding pairs between Cu and other 3d transition metals 
increased in the Cu-ordered phases (cf. Table~\ref{table.ap}).
\subsection{Magnetic ordering}
\label{subsec.mag}
\begin{figure*}[htbp]
  \begin{center}
    \includegraphics[width=1.0\linewidth]{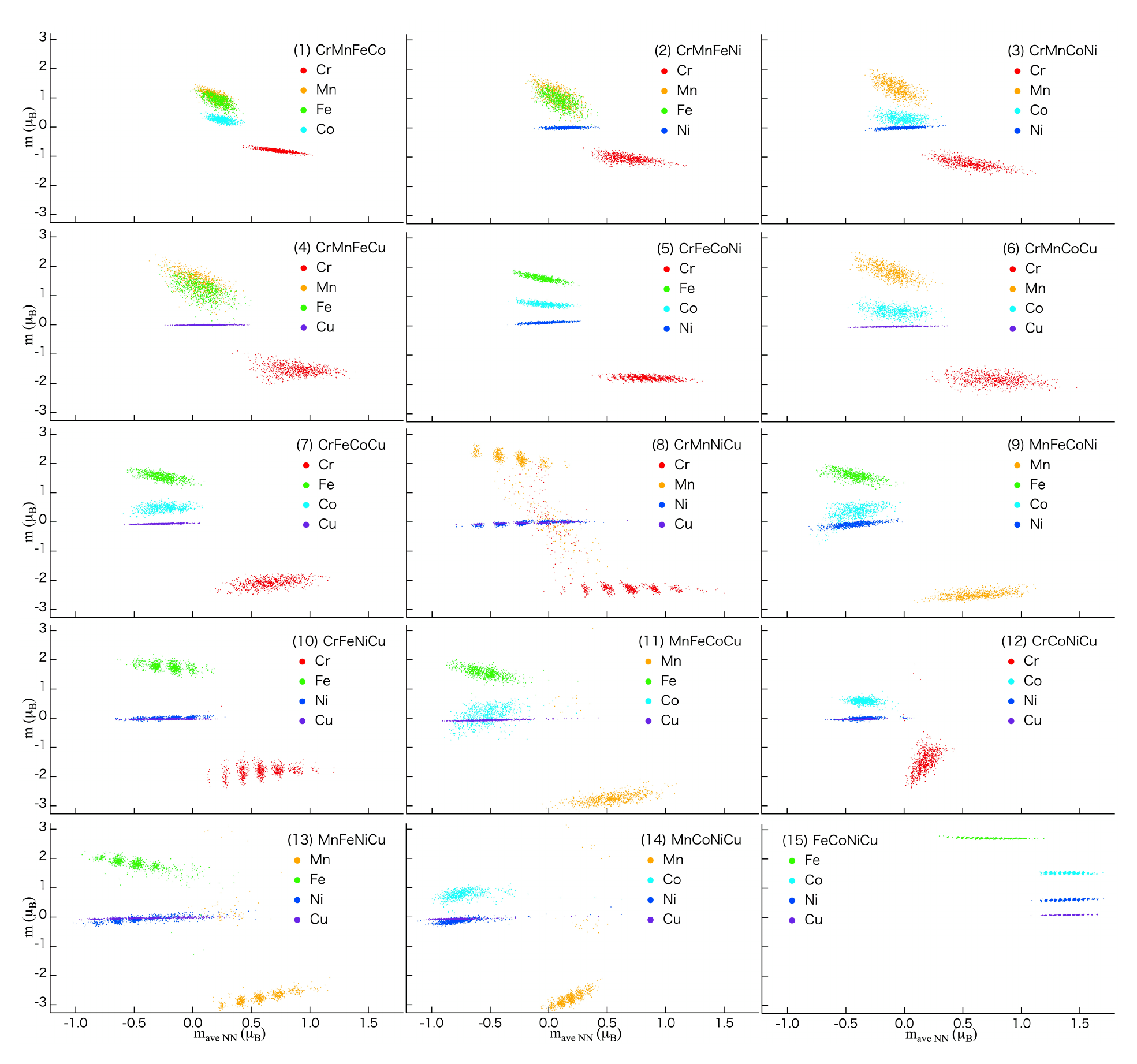}
    \caption{
      Distribution of magnetic moments as a function of
      the average magnetic moment at the first nearest-neighbor site
      in the most energetically stable ordered phase
      for the 15 quaternary alloys
      listed in Table~\ref{table.main} as OP$_0$.
      Red, orange, green, light blue, dark blue, and purple indicate
      Cr, Mn, Fe, Co, Ni, and Cu, respectively.
      $m$ and $m_{\mathrm{ave~\!\!NN}}$ represent
      the magnetic moment for each atom and the average magnetic moment
      of the 12 first nearest-neighbor sites, respectively.
    }
    \label{figure.mm}
  \end{center}  
\end{figure*}

Figure~\ref{figure.mm} illustrates the distribution of magnetic moments
as a function of the average magnetic moment at the first nearest-neighbor sites
in the most energetically stable ordered phase for 15 quaternary alloys.
In Fig.~\ref{figure.mm}, we observe certain common trends: 
``Cr'' and ``Mn without Cr'' show antiferromagnetic properties, 
``Mn with Cr'', ``Fe'', and ``Co'' show ferromagnetic properties, 
and ``Mn with a ferromagnetic moment'' shows values that can be compared to those of Fe.
In contrast, Mn with antiferromagnetic properties behave similar to Cr.
Therefore, the magnetic nature of Mn depends on the presence or absence of Cr.
Ni and Cu exhibit nonmagnetic (paramagnetic) behavior in all systems.
Although the magnetic moments are mostly similar, the magnetic moment distributions of some alloys 
show that the average magnetic moments at the first nearest-neighbor sites are discrete; 
for example, at Cr and Fe in CrFeNiCu (VEC$_{\mathrm{ave}} =$8.75), as shown in Fig.~\ref{figure.fe} (10). 
Such a discrete distribution can be attributed 
to the number of ferromagnetic or antiferromagnetic atoms 
located at the first nearest-neighbor sites of the FCC lattice (Fig. 2(b) in~\cite{2015NIU}).
For example, the antiferromagnetic atoms in the ordered phases 
are surrounded by four ferromagnetic atoms on average 
because the other three elements are occupied at the 12 first nearest-neighbor sites.

Niu et al.~\cite{2015NIU}
previously evaluated the magnetic moment of each element 
and the average magnetic moment of the 12 first-nearest neighbor sites
for the RSS and Cr-\Lot{} ordered phases of the CrFeCoNi quaternary alloy 
(VEC$_{\mathrm{ave}} = 8.25$), and they are the same as that of the composite shown in Fig.~\ref{figure.fe} (5).
The results obtained in this study demonstrate that 
Fe and Co exhibit ferromagnetic properties, whereas Cr exhibits anti-ferromagnetic properties.
Since the results obtained are consistent with those reported previously ~\cite{2015NIU}, 
the validity of the calculation methods described herein can be confirmed.
Moreover, Niu et al. reported a discrete distribution of the average magnetic moments 
at the first nearest-neighbor sites in the Cr-\Lot{} ordered phase.
They revealed that
the magnetism of Cr in the CrFeCoNi quaternary alloy
is strongly influenced by the magnetism of the first nearest-neighbor sites~\cite{2015NIU}.
 Zuo et al.~\cite{2017ZUO} evaluated the magnetic moments of the individual atoms
 of the FCC MnFeCoNi quaternary alloy and FCC CrMnFeCoNi quinary alloy 
and reported that Fe exhibits a higher magnetic moment than Co, 
Ni exhibits a near-zero magnetic moment,
and Cr exhibits antiferromagnetism. 
In addition, Mn atoms cause a wider distribution in the magnetic moments of atoms 
and exhibit both ferromagnetism and antiferromagnetism in the RSS phase.~\cite{2017ZUO}

Ghazisaeidi et al. reported 
a similar magnetism distribution in the hexagonal close packed (HCP)
and FCC phases of CrCoNi, FeCoNi, MnFeNi, MnCoNi, and CrMnFeCoNi.~\cite{2018NIU} 
The magnetism of Cr and Mn atoms
ranges from ferromagnetism to antiferromagnetism in the RSS phase.
Fedorov et al.~\cite{2020FED} and Sch\"{o}nfeld et al.~\cite{2019SCH}
out that the magnetic properties of the constituent elements
are key factors governing phase stability and
the formation of ordered phases in 3d-transition-metal-based alloys.

\subsection{Quantitative Structure--Property Relationship}
\label{subsec.rel}
The \Lot{} and \Dzt{} phases have
 typical ordered crystal structures in the 25:75 (at.\%) FCC binary alloys.
For example, the \Lot{} structure can be observed in many alloys with a 3:1 composition, such as Cu$_3$Au, Ni$_3$Al, and TiPt$_3$; similarly, the \Dzt{} structure
can be observed in trialuminides Al$_3$ (Sc, Ti, V) and Ni$_3$V.
Although the appearance of the \Lot{} and \Dzt{} phases in FCC binary alloys
has been discussed previously,
only a few studies have focused on the configuration of the neighboring sites
in multiprincipal-element alloys.~\cite{2015NIU}
The number of first, second, third, and fourth nearest-neighbor sites in the \Lot{} and \Dzt{} phases
are 12, 6, 24, and 12, respectively (Table~\ref{table.ap}). 
The 25 at.\% element in the \Lot{} and \Dzt{} phases is
not located at the first nearest neighbor sites, and thus,
 they can be isolated from each other and
surrounded by the 75 at.\% elements.
For instance, Au atoms in the \Lot{} phase of Cu$_3$Au are
not located at their first nearest-neighbor sites and are surrounded by 12 Cu atoms.
\begin{table}
 \begin{center}
   \caption{
     Comparison of the site occupancies     
     of \Lot{}, \Dzt{}, and RSS phases containing 25 at.\% of ordered elements.
     For each neighbor site therein, the number of the ordered elements over
     the total number of the sites is given with its percentage in brackets.
     For the RSS phase, the occupancy rate for any site is 25 at.\% on average (shown in brackets).     
     } 
     \label{table.ap}
     \begin{tabular}{crrr} 
       Neighbor site & \multicolumn{1}{c}{\Lot{}}  & \multicolumn{1}{c}{\Dzt{}} & \multicolumn{1}{c}{RSS} \\
       \hline
       First nearest & 0/12 $\langle 0\rangle$ & 0/12 $\langle 0\rangle$ & $\langle 25 \rangle$\\ 
       Second nearest & 6/6 $\langle 100\rangle$ & 4/6 $\langle 67\rangle$ & $\langle 25 \rangle$\\ 
       Third nearest & 0/24 $\langle 0\rangle$ & 8/24 $\langle 33\rangle$ & $\langle 25 \rangle$\\ 
       Fourth nearest & 12/12 $\langle 100\rangle$ & 4/12 $\langle 33\rangle$ & $\langle 25 \rangle$\\ 
       \hline
     \end{tabular}
 \end{center}
\end{table}

\begin{figure}[htbp]
  \begin{center}
    \includegraphics[width=0.9\linewidth]{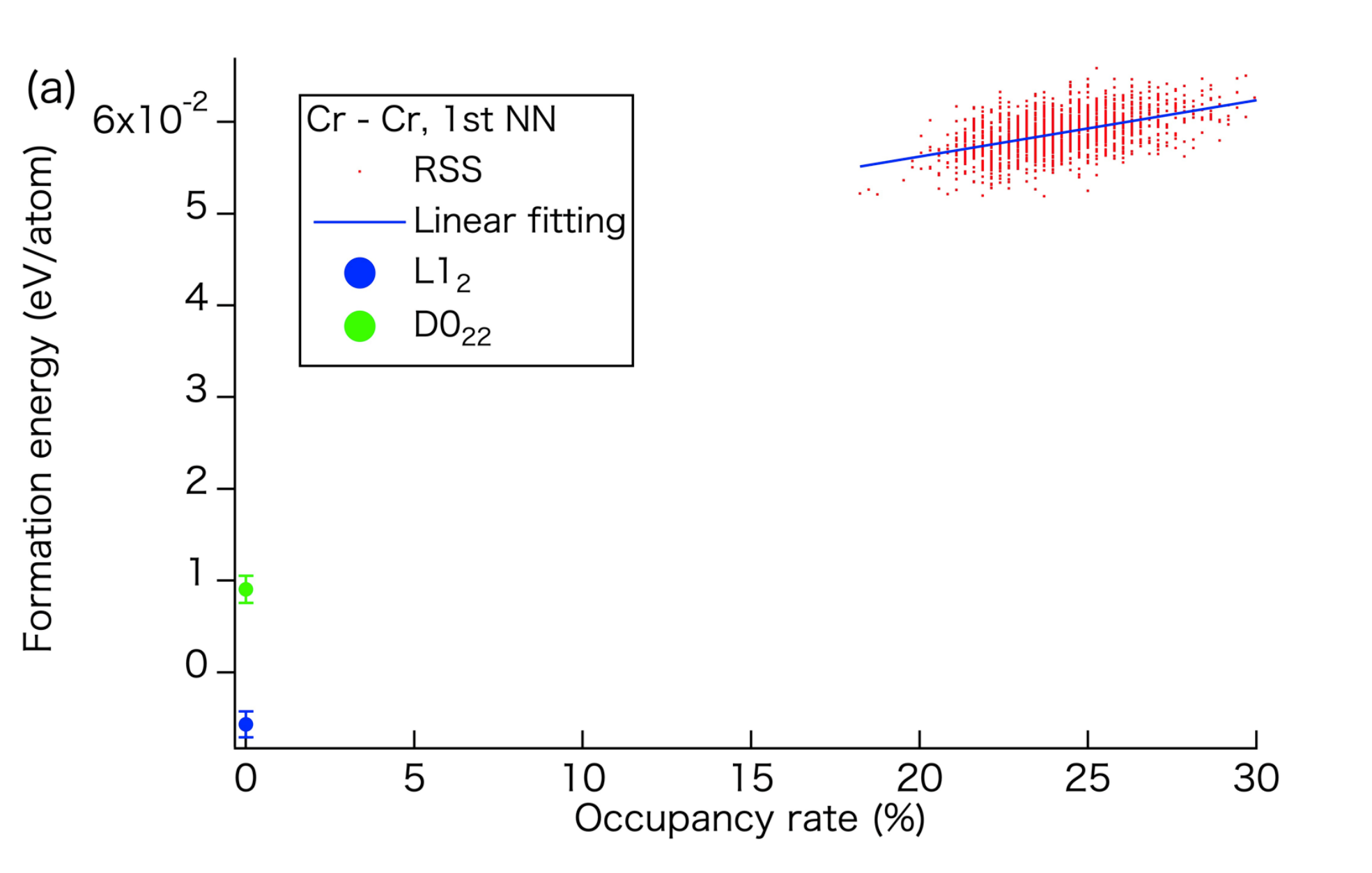}
    \includegraphics[width=0.9\linewidth]{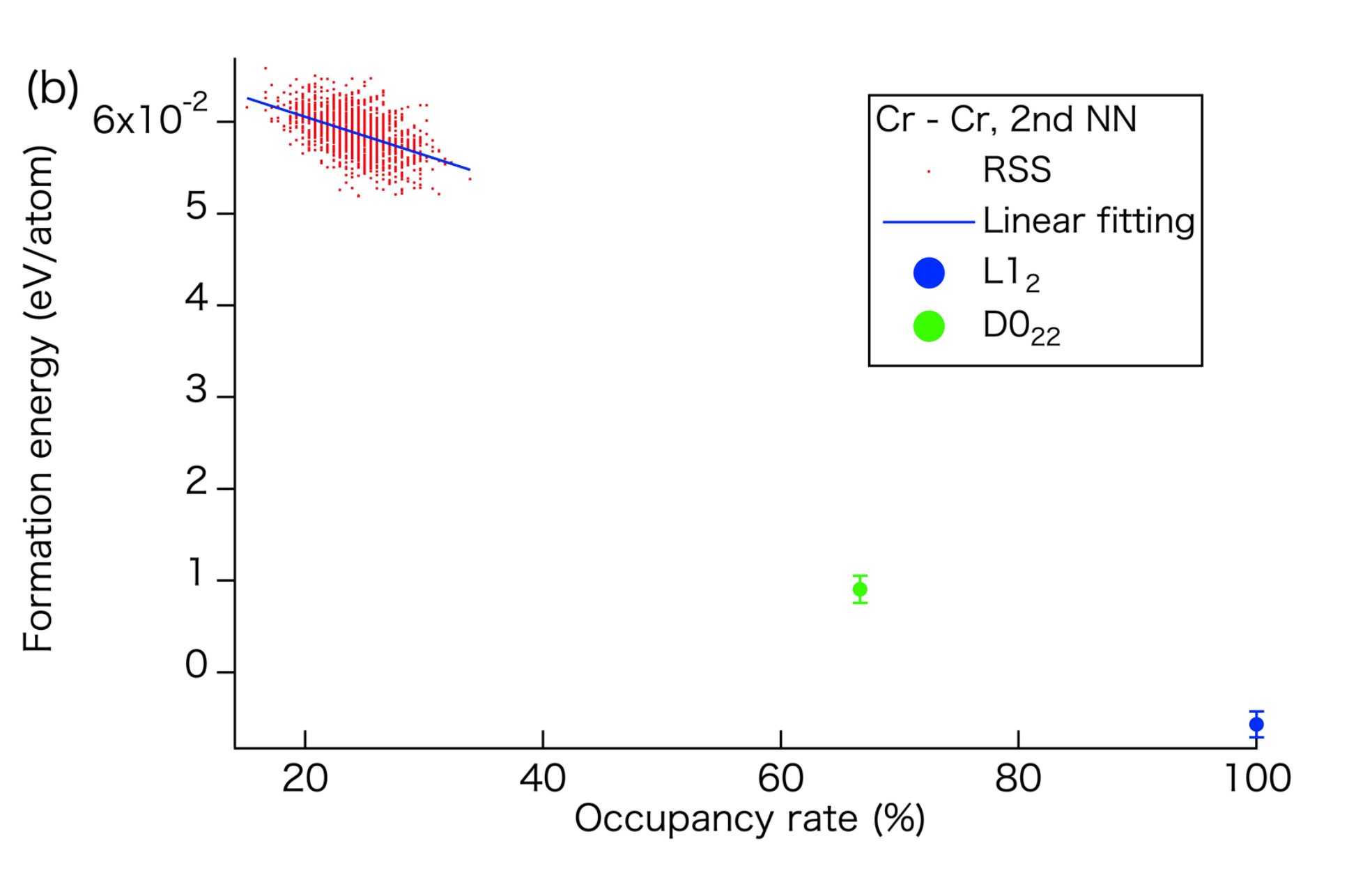}        
    \caption{
      Dependence of the formation energies of ordered and RSS phases 
      in the CrFeCoNi quaternary alloy on the occupancy rate of the Cr--Cr pairs 
      at the (a) first and (b) second nearest-neighbor sites. 
      The red dots represent the RSS phase obtained using 1,500 samples,
      whereas the blue lines serve as guidelines.
      The blue and green dots indicate the Cr-\Lot{} and Cr-\Dzt{} ordered phases, respectively.
    }
    \label{figure.ed}
  \end{center}  
\end{figure}

Figure~\ref{figure.ed} shows the relationship 
between the formation energy and occupancy rate of Cr--Cr pairs 
at the first and second nearest-neighbor sites in the RSS phase of the CrFeCoNi quaternary alloy. 
In the RSS phase, 
the rate of occupancy by the same element can remain at 25 at.\% on average 
at any nearest-neighbor site. 
Meanwhile, the occupancy rates for the ordered phases 
depend on the nearest-neighbor site, as listed in TABLE~\ref{table.ap}. 
For the first nearest-neighbor sites (Fig.~\ref{figure.ed}(a)), 
the calculated results reveal a lowering of the energy with a decreasing occupancy rate, 
such that the energy approaches 
the values corresponding to the \Lot{} and \Dzt{} ordered phases on extrapolation of the occupancy rate to zero. 
However, for the second nearest-neighbor sites (Fig.~\ref{figure.ed}(b)), 
an opposite trend is observed, i.e., 
the formation energies decrease with increasing occupancy rate. 
The values for both ordered phases can be found on the extrapolation line. 
These results imply that the configurations at the first and second nearest-neighbor sites 
affect the formation energy of the CrFeCoNi quaternary alloy. 
Furthermore, the lower formation energies of the ordered phases 
can be attributed to their unique occupancy rates. 

\begin{figure}[htbp]
  \begin{center}
    \includegraphics[width=1.0\linewidth]{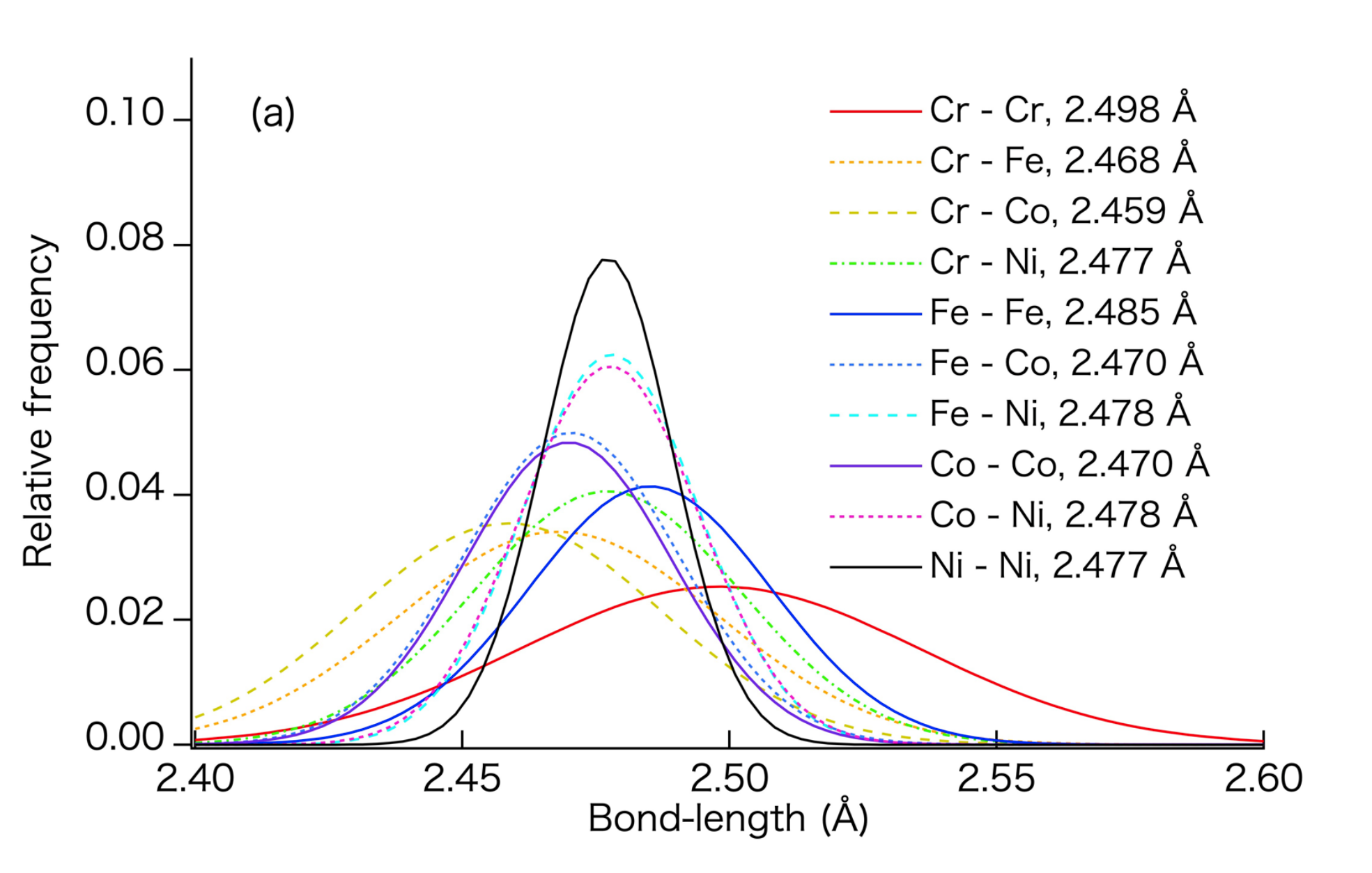}
    \includegraphics[width=1.0\linewidth]{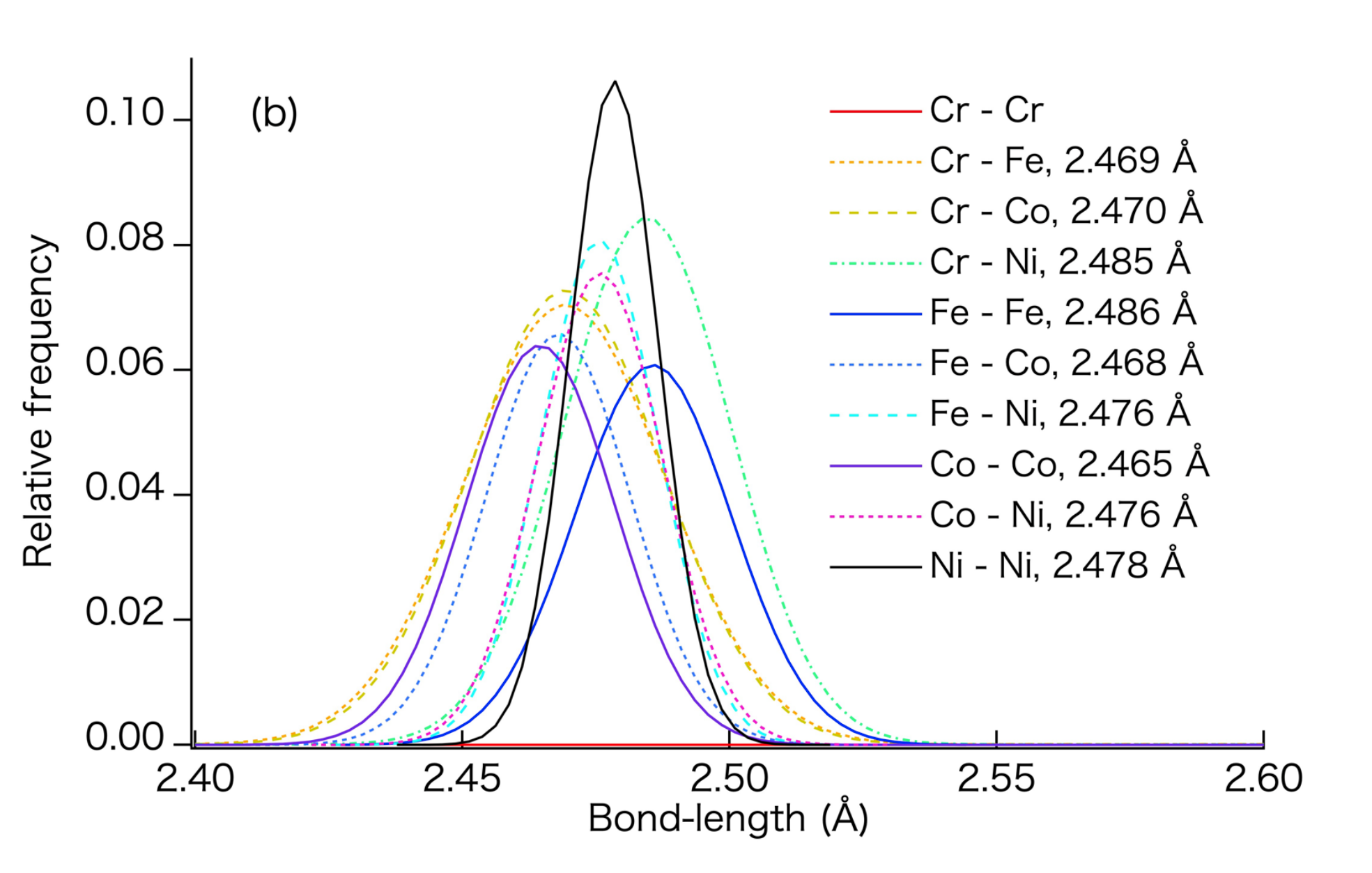}    
    \caption{
      Relative frequency distribution of bond-lengths 
      fitted with Gaussian function for each element pair in the 
      (a) RSS phase and (b) Cr-\Lot{} ordered phase of the 
      CrFeCoNi quaternary alloy after structure optimization.
      The bond lengths in the legends indicate the average values
      obtained upon fitting with the Gaussian function.
    }
    \label{figure.rdf}
  \end{center}  
\end{figure}

Figures~\ref{figure.rdf} (a) and (b) show
the bond-length distributions in 
the RSS and Cr-\Lot{} ordered phases, respectively.
In Figure~\ref{figure.rdf},
the solid lines indicate homogeneous element pairs, whereas 
the dotted and dashed lines indicate heterogeneous element pairs. 
Many 3d transition metals, especially Cr, Fe, Co, and Ni, 
exhibit comparable atomic radii. 
However, the CrFeCoNi quaternary alloy 
shows a different bond-length distribution. 
These results indicate that these element pairs 
have different interatomic distances. 
As shown in Figure~\ref{figure.rdf}(a), the Cr--Cr pairs in the RSS phase 
exhibit a broad bond-length distribution and longer bond lengths
than those corresponding to the first nearest-neighbor sites, as estimated by the lattice constant.
The bond lengths of the homogeneous element pairs 
are in a decreasing order of Cr, Fe, Ni, and Co.
The same trend is observed for the Cr-\Lot{} ordered phase in Fig.~\ref{figure.rdf}(b). 
Notably, there are no Cr--Cr pairs in the Cr-\Lot{} ordered phase 
because Cr atoms occupy the sites constructing 
the framework of the \Lot{} ordered phase, and they
 are not located at the first nearest-neighbor sites (TABLE~\ref{table.ap}).
Although the structure optimization process is applied to
all atoms in the Cr-\Lot{} ordered and RSS phases, 
the bond-length distributions of the Cr-\Lot{} ordered phase 
are narrower than those of the RSS phase.

Let us consider the half widths at half maximum (HWHM) of the fitted Gaussian functions.
The HWHMs of Fe--Fe, Co--Co, and Ni--Ni in the \Lot{} ordered phases are
64.5 \%, 71.2 \%, and 68.6 \%, respectively, of the corresponding RSS phases.
It is presumed that the Cr occupancy at the second nearest-neighbor sites 
results in a small bond-length distribution with regularity. 
Gao and Alman used ab initio molecular dynamics (AIMD) simulations~\cite{2013GAO} and reported that the Al$_{1.3}$CrFeCoNiCu senary alloy 
exhibits a different distribution for each element pair, 
which is similar to the results obtained in this study 
for 3d-transition-metal-based equiatomic quaternary alloys. 
However, the HfNbTaTiZr quinary alloys 
show almost the same distribution for all its constituent elements without a 3d transition metal.~\cite{2013GAO} 
In this study, we conclude that the bond-length distributions 
of the described systems 
are significantly different 
because each 3d transition metal has an individual magnetic behavior.

\section{Summary and Future Prospects}
\label{sec.conc}

In this study, we performed first-principles calculations considering the configurational entropy term
to investigate the effect of VEC and temperature
on the stabilities of the two ordered phases: \Lot{} and \Dzt{}.
Further, we drew a comparison with the disordered atomic configurations 
in equiatomic quaternary alloys comprising Cr, Mn, Fe, Co, Ni, and Cu.
We found that the VEC is a significant factor
controlling phase stability in both ordered phases. 
In the high-VEC region, the \Dzt{} ordered phase
is energetically more stable than the \Lot{} ordered phase,
though both phases are  metastable.
The magnetic moments indicate
that the anti-ferromagnetic Cr and Mn atoms 
are surrounded by the ferromagnetic Fe, Co, and Ni atoms
located at the first nearest-neighbor sites. 
Such magnetic configurations make the ordered HEA phases energetically stable.
Since the constituent elements corresponding to 25 at.\%
in the \Lot{} and \Dzt{} ordered phases
exhibit no bonding between the first nearest-neighbor sites,
the extent of the combination of anti-ferromagnetic
and ferromagnetic elements can be increased in the \Lot{} and \Dzt{} phases.
The bond-length distribution 
results reveal that the Cr--Cr distances are longer 
and the distributions are broader than those of other element pairs, 
which suggests that the high formation energy of the RSS phase
can be attributed to the abnormal bond length
between the anti-ferromagnetic atoms. 
Consequently, both the ordered phases have
enthalpy-driven energetically favorable structures that are enthalpy-driven.

\par
Since 3d-transition-metal quaternary systems are base alloys
for their extension to more multicomponent systems,
a comprehensive understanding of their structures would serve as a stepping stone toward
efficient HEA design.
As shown for most 3d-transition-metal quaternary alloys, 
we should consider that phase stability is realized by 
an order–-disorder ``competitive cooperation,''
rather than a usual order–-disorder competition.
The former was newly proposed in this study, 
while the latter has already been accepted in the HEA community.
Most researchers in the HEA community naively consider that 
the ``fully disordered/random atomic configurations'' of HEAs 
have higher entropies, and therefore, they have higher phase stabilities. 
However, it is important to recognize the cooperation between
``a magnetically ordered atomic configuration composed of one element''
and ``fully disordered/random configurations composed of the other three elements,''
which correspond to the enthalpy and entropy terms, respectively, to understand the microscopic structures of HEAs in detail.

\par
In this study, the enthalpy terms that describe structural properties
were evaluated by first-principles calculations
to obtain insight into an atomistic origin of the energetics of the 3d-transition-element HEAs.
Since the entropy term is as important as the enthalpy term,
the addition of a configurational entropy term in the density functional theory (DFT) calculations 
is effective for determining the crystal structure of HEAs.
This enables us to estimate the transition temperatures between the ordered and RSS phases,
which is useful for obtaining the relationship between structure and processing.
It is already widely recognized in the materials science field
that the relationship among structures, properties,
and processing methods is a crucial factor in realizing outstanding
structural materials~\cite{2016XIO}. 
For example, Kenel et al. designed 
a 3d-transition-metal HEA using a 3D printing technique~\cite{2020HAN2} with oxide nanopowder,
which exhibits prominent
mechanical properties at ambient and cryogenic temperatures~\cite{2019KEN}.
Futhermore, it is necessary to predict an accurate phase diagram of HEA
to realize QSPR by combining 
the calculation of the phase diagram (CALPHAD)~\cite{1998SAU}
and the first-principles based thermodynamic assessment~\cite{2020HAN1}.
We believe that the systematic investigation described herein
will be helpful for realizing such an assessment and for predicting the microstructures of HEA
by performing model simulations such as ``first-principles phase field model''
without any thermodynamic empirical parameters~\cite{2019BHA}.

\par
Finally, we note that our finding is applicable to 
``ceramics'', ``oxides,'' and ``carbide'', as well because 
the ``high-entropy'' concept has been adopted 
even in them~\cite{2018ANA,2019SAR,2019HAR,2019YE,2020OSE,2020KAU,2021MCC,2021HOS}.
Furthermore, 
our proposed concept sheds light on the constituent phases of even multiprincipal element alloys (MPEAs)~\cite{2015SEN,2019ZHE}
and multiphase compositionally complex alloys (CCAs)~\cite{2019MAN}.
Thus, our finding can encourage experimentalists to 
investigate the microscopic identification of their structures,
which will help us elucidate their QSPR.


\setcounter{figure}{1}
\renewcommand{\thetable}{S-\Roman{table}}
\renewcommand{\thefigure}{S-\arabic{figure}}
\section{Supplementary Information}
\subsection{{\it Ab initio} calculations}
\label{computational}
Figure~S-1 shows the formation free energy
as a function of temperature
for 15 kinds of quaternary alloys, 
which are listed at Table I. 
The slope of the formation free energy for RSS phase 
is steeper than that of the ordered phases, 
because the entropy term of RSS is larger than that of the ordered phases. 
On the other hand, the slope of both ordered phases are exact same. 
Consequently, the comparison of the formation free energy 
of the ordered phases is determined by formation energy at 0 K. 
Note that in the case of CrFeCoCu (VEC$_{\mathrm{ave}} = 8.50$) alloy in (7) at Fig.~S-1, 
the both ordered phases have almost same formation free energy. 
Figure~S-1 (14) and Figure~2 show the formation free energy for MnCoNiCu alloy 
with different vertical scale. 
In Fig.~S-1, different scale is set for comparison with other alloy systems.

Figure~S-2 illustrates the distribution 
of the formation free energies of at 1,000 K for quaternary alloys
using 100 samples.
Figure 3 shows the average of formation free energies
with the standard deviation from the raw data of 100 samples
in Fig.~S-2. 
$X$-\Lot{} and $X$-\Dzt{} ($X =$ Cr, Mn, Fe, Co, Ni, and Cu) 
indicate the cubic corner (CC) sites of an face centered cubic (FCC) lattice 
occupied by each element for \Lot{} structure
and the corner sites and body centered sites 
of an FCC lattice when $c = 2a$ for \Dzt{} structure, respectively.
a is a lattice constant of FCC lattice.
RSS is random solid solution (cf. Figure~1).
Namely, 8 kinds of the ordered phases and RSS phase
for each composite can be compared in Fig.~S-1.
It becomes clear that the quaternary alloys
containing Cu show high formation free energy
than those of other alloys,
as a result of comparison of 15 kinds of the alloys
with the same energy ranges on all vertical axes.
A few ordered phases, such as Cr-\Lot{} phase,
show a narrow dispersion of formation free energy,
while many ordered phases show a same wide dispersion
as RSS phases. Interestingly, $X$-\Lot{} ($X =$ Cr and Mn)
phases of CrMnNiCu (VEC$_{\mathrm{ave}} = 8.50$) in (8), 
CrFeNiCu (VEC$_{\mathrm{ave}} = 8.75$) in (10), 
MnFeCoCu (VEC$_{\mathrm{ave}} = 8.75$) in (11), 
CrCoNiCu (VEC$_{\mathrm{ave}} = 9.0$) in (12), 
MnFeNiCu (VEC$_{\mathrm{ave}} = 9.0$) in (13), 
and MnCoNiCu (VEC$_{\mathrm{ave}} = 9.25$)
in (14) at Fig.~S-2 show a scatter distribution.

\medskip
\textbf{Acknowledgements} \par 
This study was supported by
the computational resources of the HPCI system
(Project ID: hp190014, hp190059, hp200040, hp210019),
the JHPCN system (Project ID: jh190019),
Institute for Materials Research, Tohoku University 
(Proposal No. 19S0501, 20S0505, 202012-SCKXX-0506),
JAIST, and NIMS. 
H.M. thanks the Korea Institute of Science and Technology 
(Grant Nos. 2Z05840, No. 2E31201, No. 2N62110)
for providing the financial support for this research.
R.S. acknowledges the support from the Council for Science,
Technology and Innovation
under the Cross-ministerial Strategic Innovation Promotion Program. 
K.H. acknowledges the financial support from 
MEXT-KAKENHI (Grant Nos. JP19K05029, JP19H05169)
and the Air Force Office of Scientific Research
(Award Number: FA2386-20-1-4036).

\medskip

%




\begin{figure}
\textbf{Table of Contents}\\

\medskip
  \includegraphics[width=5.5cm]{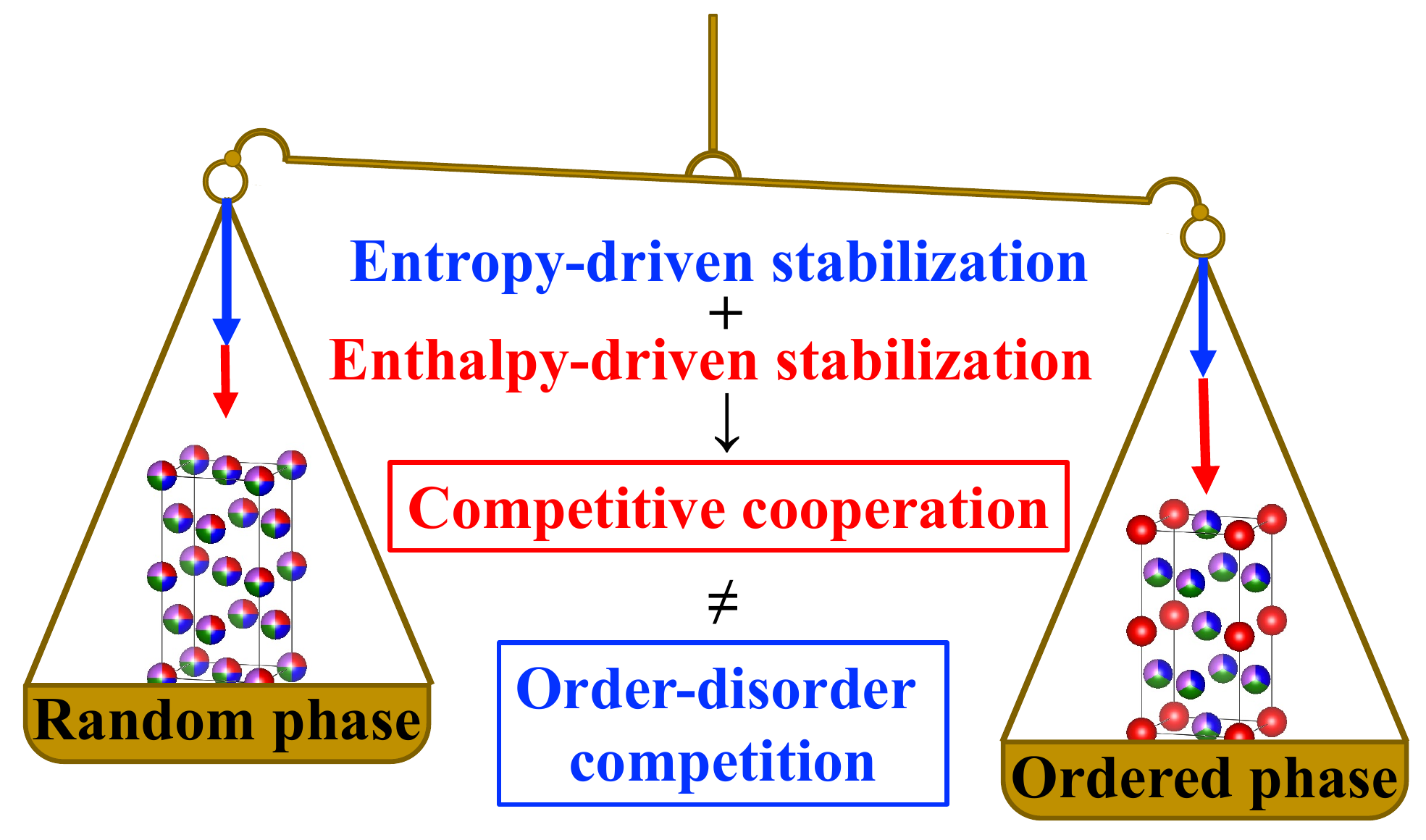}
  \medskip
  \caption*{Phase stability of 3d-transition metal
    high-entropy alloys (HEAs) is realized by 
    an order–-disorder ``competitive cooperation,''
    rather than a usual order–-disorder competition.
    The former was newly proposed in this study, 
    while the latter has already been accepted in the HEA community.}
\end{figure}

\medskip
\newpage
\begin{figure*}[h]
  \begin{center}
    \ContinuedFloat
    \subfloat{\includegraphics[width=.70\linewidth]{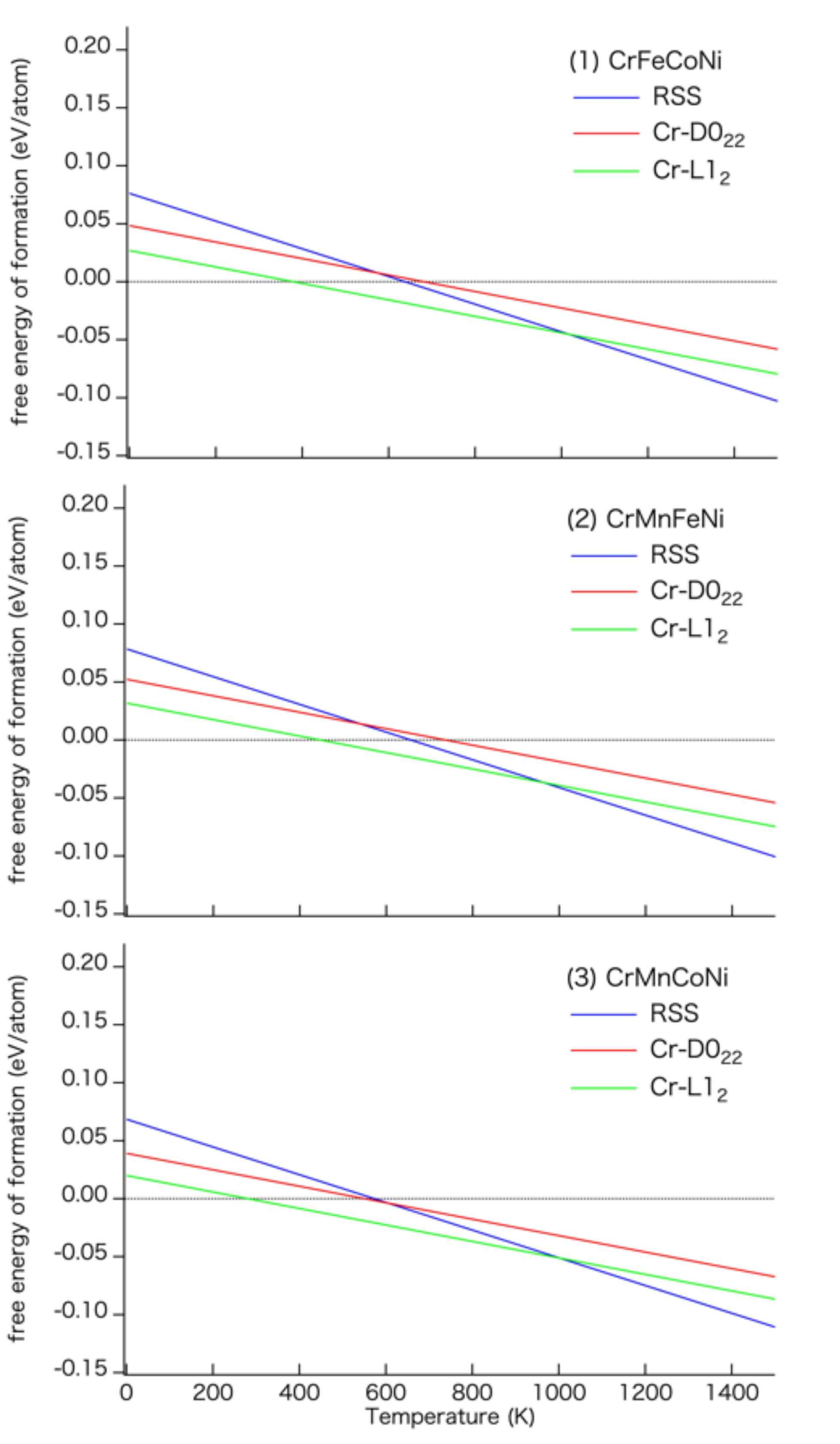}} \,    
    \phantomcaption
    \label{figure.feI}
  \end{center}  
\end{figure*}
\begin{figure*}[h]
  \begin{center}
    \ContinuedFloat
    \subfloat{\includegraphics[width=.70\linewidth]{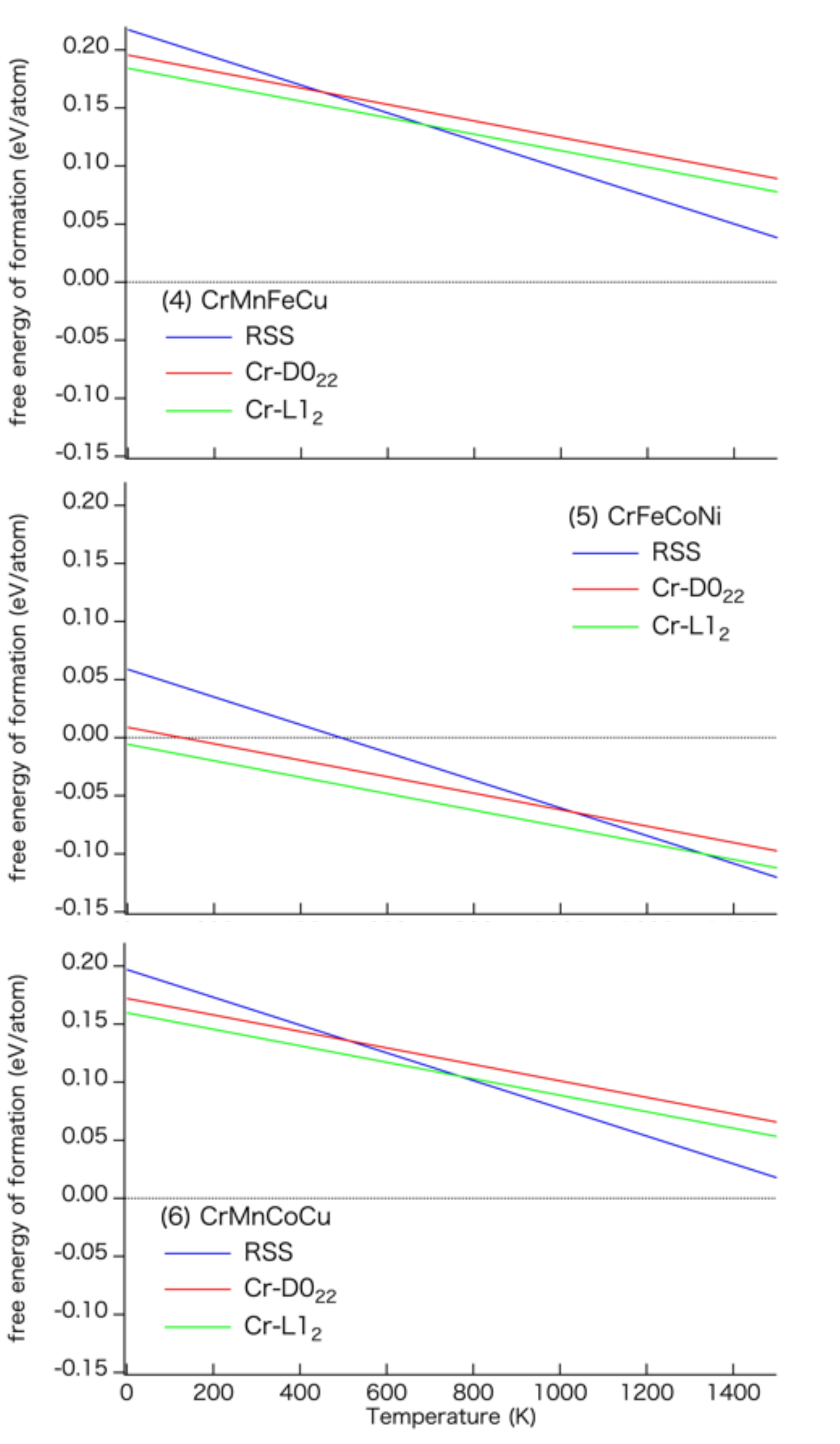}} \,
    \phantomcaption
    \label{figure.feIII}
  \end{center}  
\end{figure*}
\begin{figure*}[h]
  \begin{center}
    \ContinuedFloat
    \subfloat{\includegraphics[width=.70\linewidth]{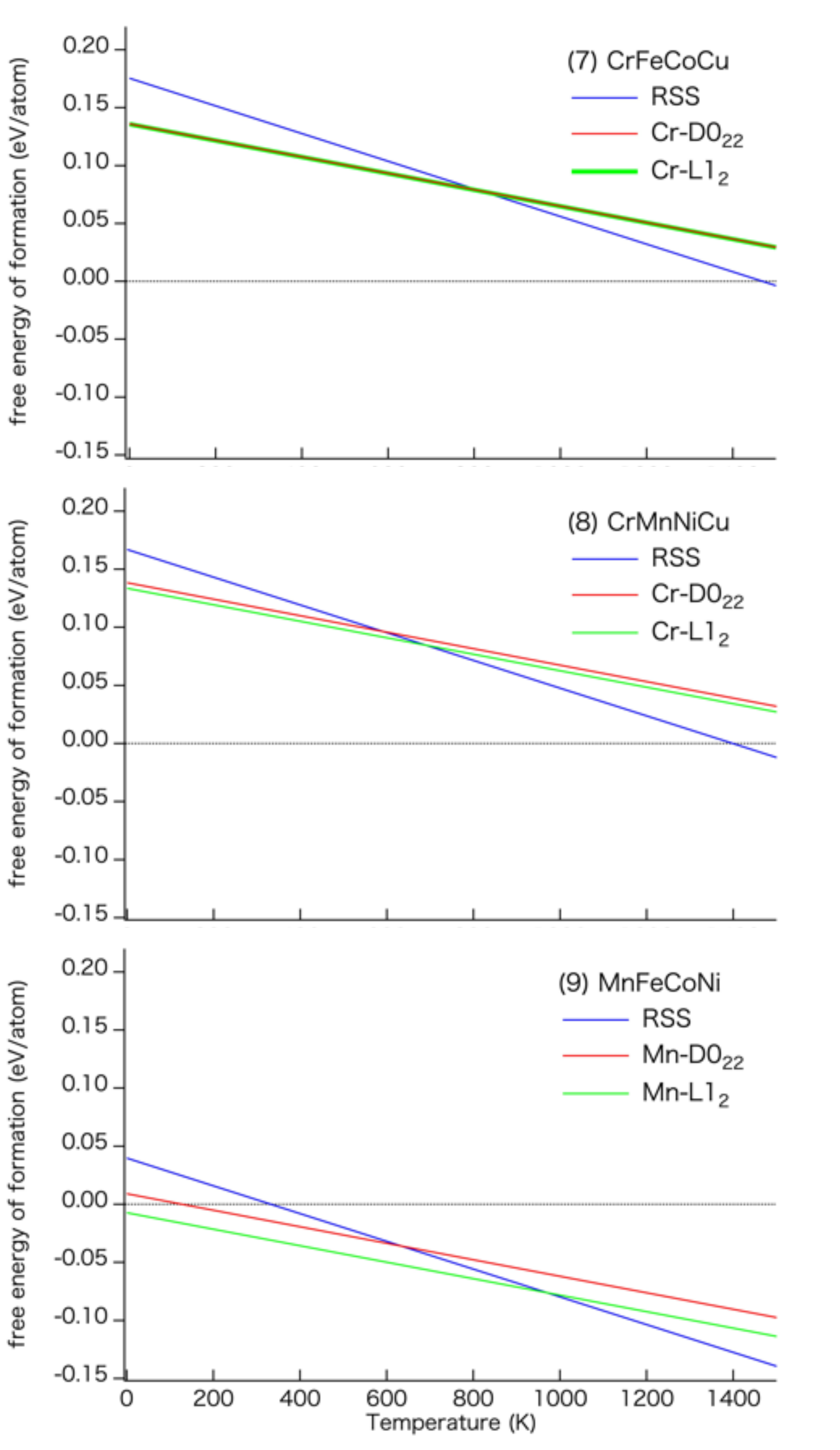}} \,
    \phantomcaption
    \label{figure.feV}
  \end{center}  
\end{figure*}
\begin{figure*}[h]
  \begin{center}
    \ContinuedFloat
    \subfloat{\includegraphics[width=.70\linewidth]{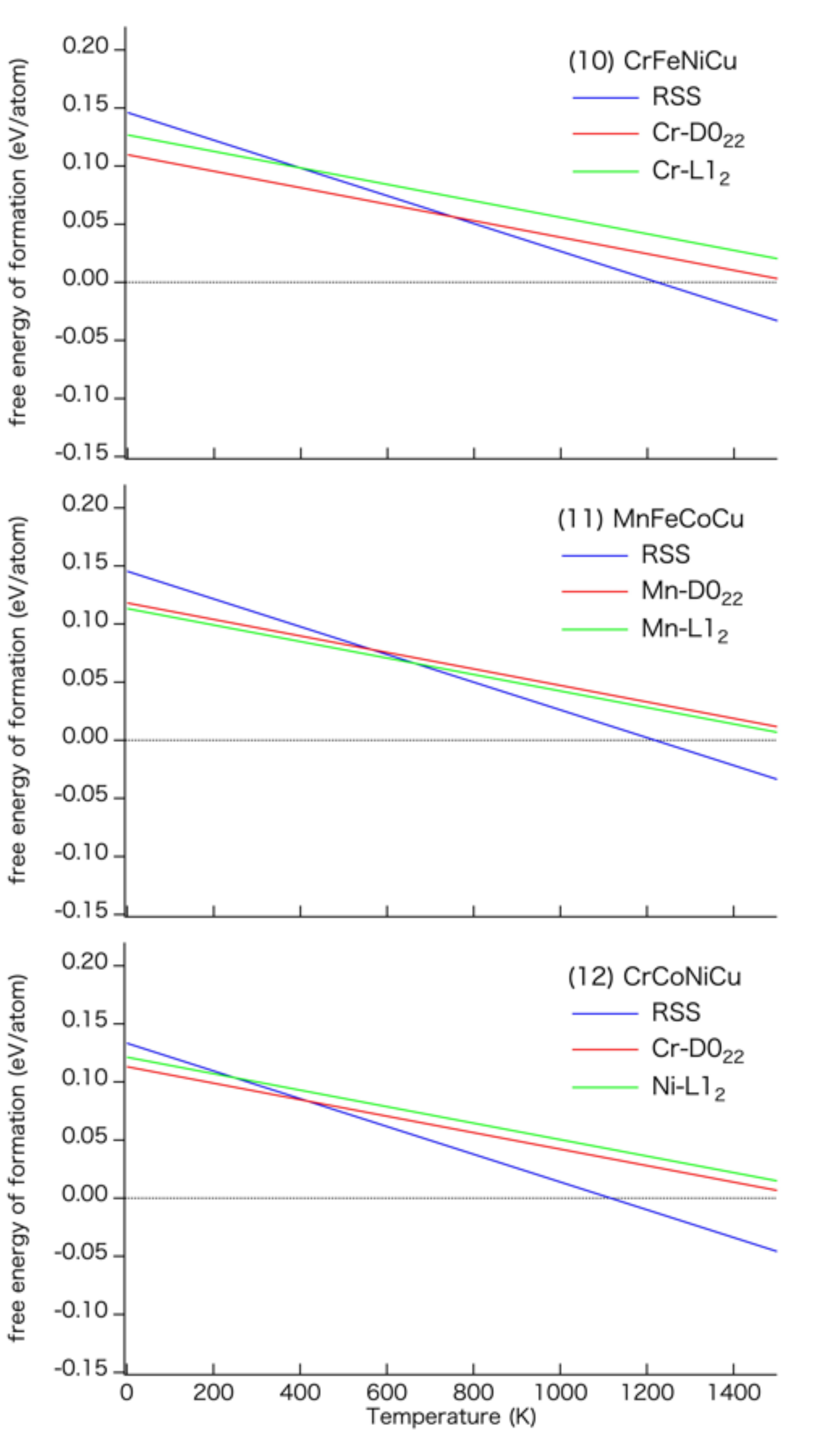}} \,
    \phantomcaption
    \label{figure.feIX}
  \end{center}  
\end{figure*}
\begin{figure*}[h]
  \begin{center}
    \ContinuedFloat
    \subfloat{\includegraphics[width=.70\linewidth]{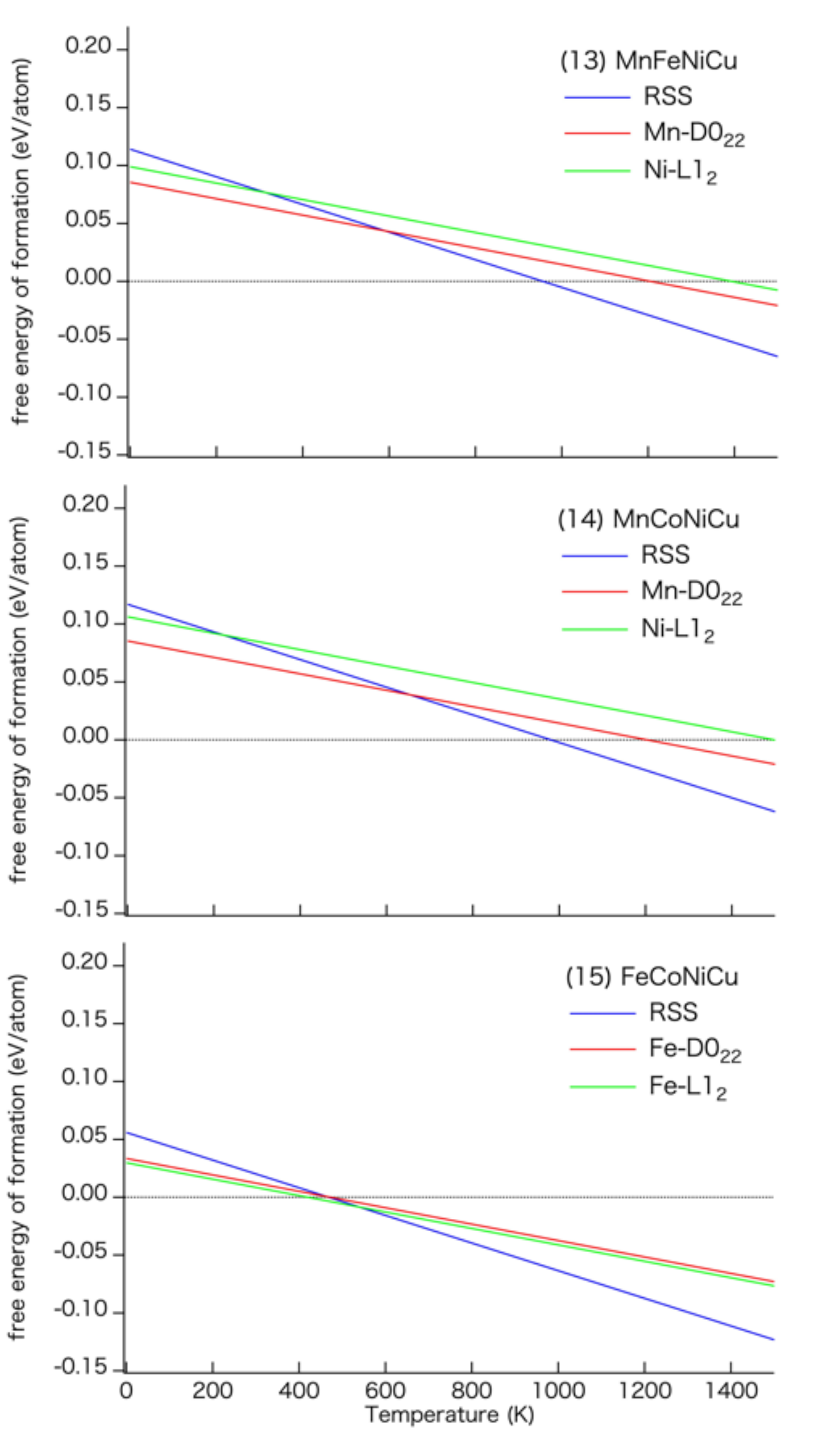}} \,
    \caption{
      Formation free energy as a function of temperature
      for 15 kinds of quaternary alloys,
      which are listed at Table I. 
      Blue, red and green lines indicate RSS, \Dzt{},
      and \Lot{} ordered phases, respectively.
    }
    \label{figure.feXV}
  \end{center}  
\end{figure*}

\setcounter{figure}{2}
\begin{figure*}[h]
  \begin{center}
    \ContinuedFloat
    \subfloat{\includegraphics[width=.70\linewidth]{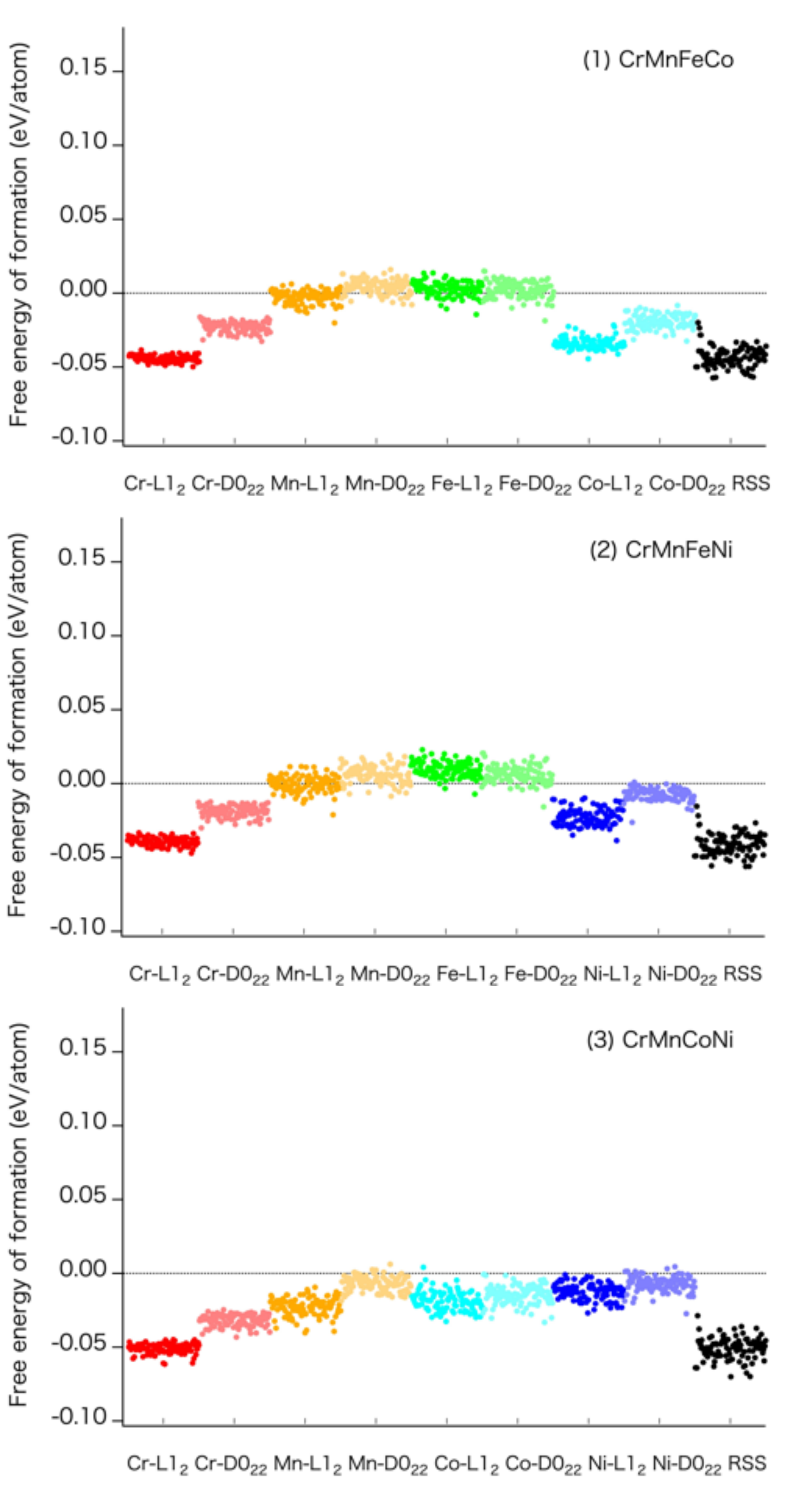}} \,
    \phantomcaption
    \label{figure.fedI}
  \end{center}  
\end{figure*}
\begin{figure*}[h]
  \begin{center}
    \ContinuedFloat
    \subfloat{\includegraphics[width=.70\linewidth]{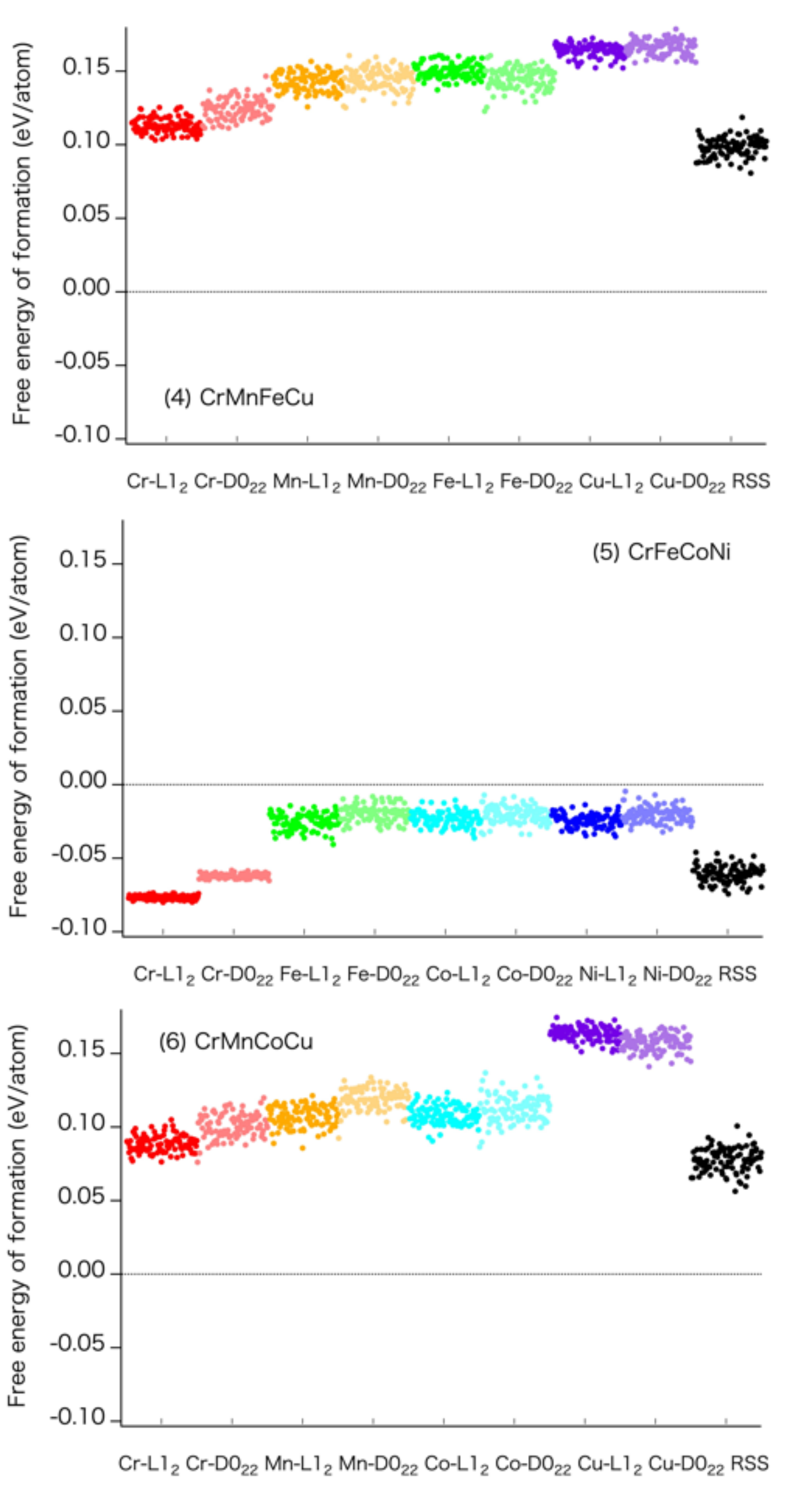}} \,
    \phantomcaption
    \label{figure.fedIV}
  \end{center}  
\end{figure*}
\begin{figure*}[h]
  \begin{center}
    \ContinuedFloat
    \subfloat{\includegraphics[width=.70\linewidth]{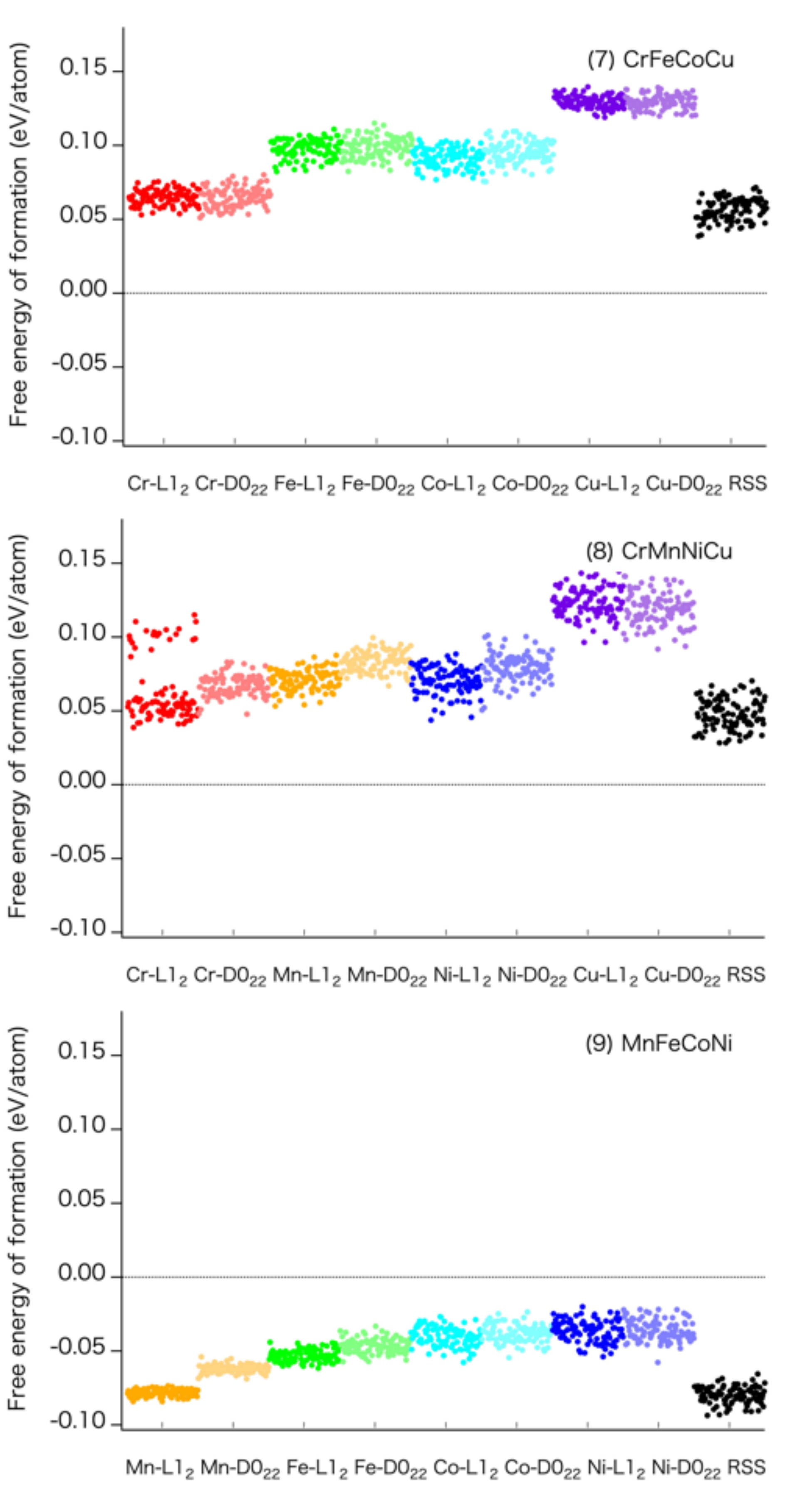}} \,
    \phantomcaption
    \label{figure.fedVII}
  \end{center}  
\end{figure*}
\begin{figure*}[h]
  \begin{center}
    \ContinuedFloat
    \subfloat{\includegraphics[width=.70\linewidth]{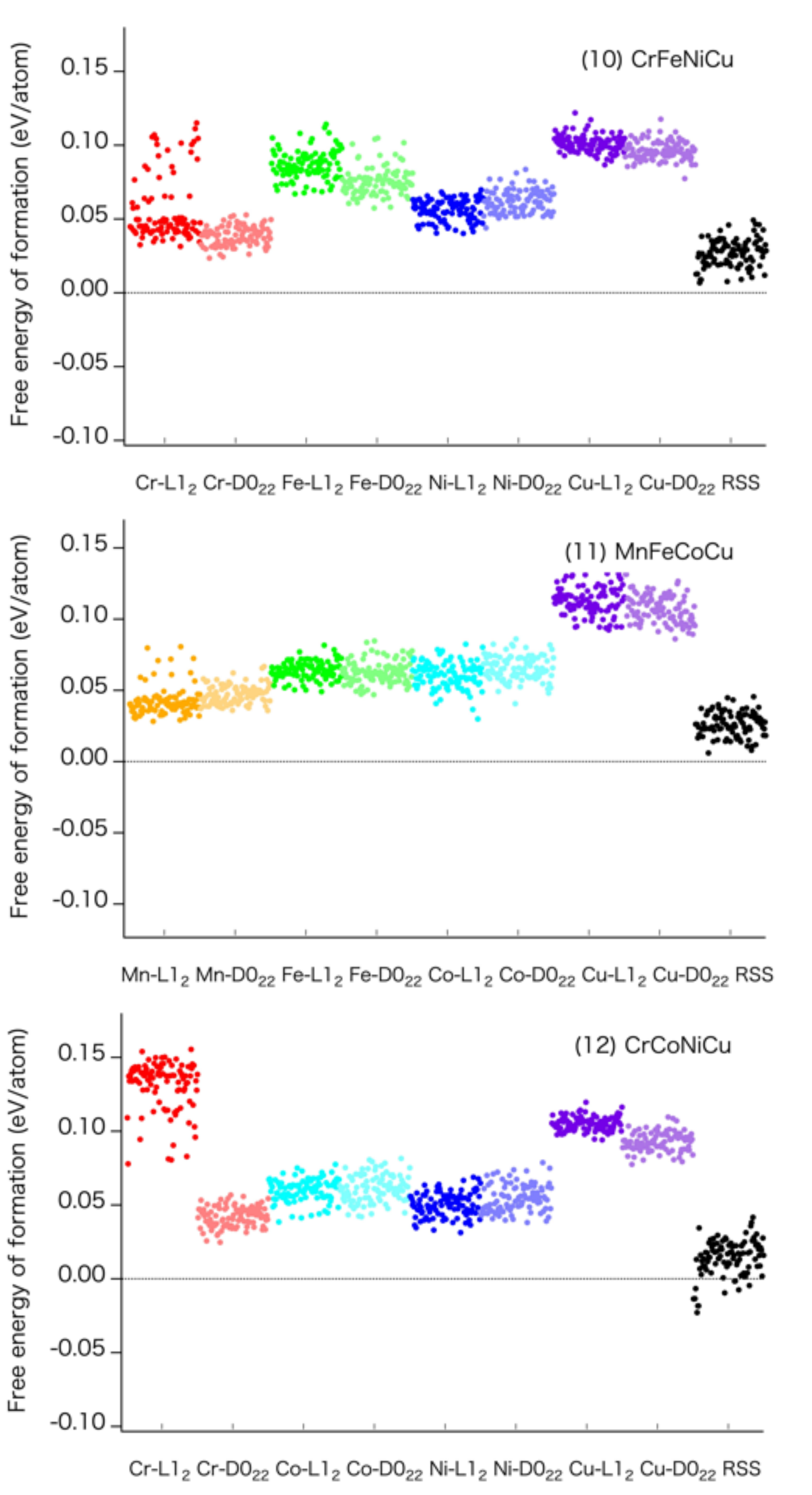}} \,
    \phantomcaption
    \label{figure.fedX}
  \end{center}  
\end{figure*}
\begin{figure*}[htbp]
  \begin{center}
    \ContinuedFloat    
    \subfloat{\includegraphics[width=.70\linewidth]{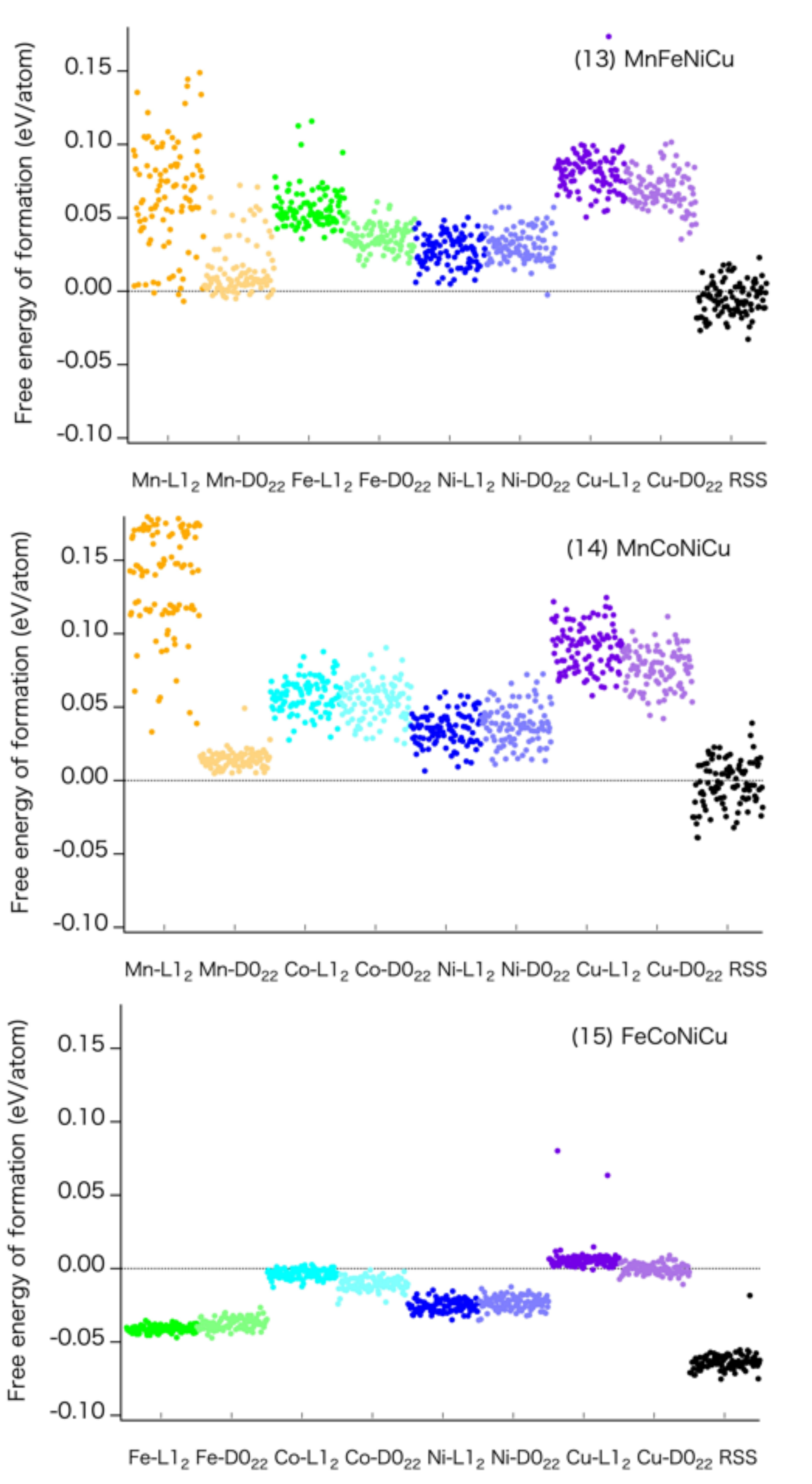}} \,
    \caption{
      Distribution of the formation free energies at 1,000 K 
      for 15 kinds of quaternary alloys,
      which are listed at Table I using 100 samples.
      Red, orange, green, cyan, blue and purple
      indicate Cr, Mn, Fe, Co, Ni, and Cu, respectively. 
      Dark and light colors indicate X-\Lot{},
      X-\Dzt{} phases, respectively. Black indicates RSS phase.
    }
    \label{figure.fedXIII}
  \end{center}  
\end{figure*}

\end{document}